\documentclass[preprintnumbers,amsmath,11pt,amssymb,floatfix,superscriptaddress,nofootinbib]{article}

\def\mytitle#1{\setcounter{equation}{0}
\setcounter{footnote}{0}
\begin{flushleft}\Large\textbf{#1}\end{flushleft}
\vspace{0.25cm}}
\def\myname#1{\leftline{{\large #1}}\vspace{-0.13cm}}
\def\myplace#1#2{\small\begin{flushleft}\textit{#1}\\
\texttt{#2}\end{flushleft}}

\def\myclassification#1{\small\noindent
Keywords :
       #1\vspace{0.5cm}}
\usepackage{graphicx}
\begin{document}
\mytitle{Thermodynamics of Reissner-Nordstr$\ddot{o}$m Black Holes in Higher Dimensions: Rainbow Gravity Background With General Uncertainty Principle }

\myname{$Amritendu~ Haldar^{*}$\footnote{amritendu.h@gmail.com} and $Ritabrata~
Biswas^{\dag}$\footnote{biswas.ritabrata@gmail.com}}
\myplace{*Department of Physics, Sripat Singh College, Jiaganj $-$ 742123, District: Murshidabad, State : West Bengal, India.\\$\dag$ Department of Mathematics, The University of Burdwan, Golapbag Academic Complex, City: Burdwan $-$ 713104, District: Purba Burdwan, State : West Bengal, India.} {}
 
\begin{abstract}
In this paper, we investigate the thermodynamic properties of Reissner-Nordstr$\ddot{o}$m black holes embedded in higher $(d)$ dimensions in the framework of rainbow gravity incorporating the effects of the generalized uncertainty principle. We also examine all the properties graphically by varying the rainbow gravity parameter $\eta$ and the generalized uncertainty principle parameter $\alpha$. We find the existence of remnant and critical mass of the concerned black hole. We calculate the local temperature, local internal energy and hence we analyse the thermal stability of the black hole by computing the local heat capacity. Further, we study the phase transitions of the aforesaid black hole solution under the effects of generalized uncertainty principle. From the analysis of the specific heat  at the horizon, we observe that there are phase transitions for all dimensions but when we analyze the same, measured by the local observer, we find that there exist only two phase transitions.
\end{abstract}

\myclassification{rainbow gravity; generalized uncertainty principle; remnant mass; local heat capacity.}\\
PACS NO: 04.60-m, 04.70.Dy, 04.70.-s\\

\section{Introduction :}

The discussions about Lorentz symmetry at Planck scale \footnote{which is assumed as an observer independent minimum measurable length scale (strongly indicated in all quantum gravity theories, viz., string theory \cite{Amati-1989}, loop quantum gravity \cite{Rovelli-1998, Carlip-2001}, Lorentzian dynamical triangulations \cite{Ambjorna-1998, Ambjornb-2000, Ambjornc-2001} and noncommutative geometry \cite{Girelli-2005} to name a few)} lead us to many possible answers. However, as the Planck scale is not a Lorentz invariant quantity, different features of observer independence of the minimum length are not Lorentz identities. Most of the theory of quantum gravity \cite{Kostelecky-1989, Hooft-1996, Amelino-1998, Gambini-1999, Carroll-2001} have proposed that the standard energy-momentum dispersion relation $c=1.$ $``E^2-p^2=m^2"$ ( $E$ is the energy of the particle, $p$ is the momentum and m is the mass of the concerned particle.) gets modified in the ultraviolet limit. One can apply this modified dispersion relation to resolve such difficulties and so appears a new theory called double special relativity \cite{Amelino-2002, Magueijo-2003, Alfaro-2002, Sahlmann-2006, Smolin-2006}. This theory is mainly based on two postulates: the constancy speed of light $c$ (from the postulate of general relativity) and the low value of Planck energy $E_p$ or Planck length $l_p$. On the basis of this new theory, the authors in the references \cite{Magueijo-2003, Magueijo-2004} have introduced rainbow gravity. In double special relativity, they have proposed that the geometry of space-time depends on  the energy of the test particle. So the geometry of space-time may be represented by a family of metrices rather than a single metric. This is why this space-time is known as rainbow gravity. The analysis of thermodynamic properties under the effects of rainbow gravity exhibits remarkably interesting results for black holes \cite{Ali-2014, Alia-2015, Ali-2015}. In rainbow gravity back ground the modified dispersion relation is stated as \cite{Magueijo-2003, Magueijo-2004}:
\begin{equation}\label{ah9.equna}
E^2 f^2 \left(\frac{E}{E_p}\right)^2- p^2 g^2 \left(\frac{E}{E_p}\right)^2= m^2~,
\end{equation}
where $p$ and $m$ are the momentum and the mass of the test particle  $f \left(\frac{E}{E_p}\right)$ and $g \left(\frac{E}{E_p}\right)$ are considered as rainbow functions. These functions play an important role to modify the standard energy-momentum dispersion relation  in the ultraviolet limit. However, these functions get constrained to reproduce the standard dispersion relation in the infrared limit, such that, $$\lim_{\frac{E}{E_p}\rightarrow 0} f \left(\frac{E}{E_p}\right)=1 ~~~ and~~~\lim_{\frac{E}{E_p}\rightarrow 0} g \left(\frac{E}{E_p}\right)=1.$$
Magueijo and Smolin \cite{Magueijob-2003} constructed modified dispersion relation according to the varying speed of light theory. This relation takes the form $\frac{E^2}{\left(1-\gamma \frac{E}{E_p}\right)^2}-p^2=m^2$ \cite{Magueijob-2003}. This modified dispersion relation signifies that space-time has an energy dependent light's speed given as : $c= \left(1-\gamma \frac{E}{E_p}\right)$, at the limit ${\frac{E}{E_p}\rightarrow 0}$, we get $c \rightarrow 1$.
Due to loop quantum gravity, introducing a rainbow gravity parameter $\eta$, these rainbow gravity functions become 
\begin{equation}
f \left(\frac{E}{E_p}\right)=1 ~~~ and ~~~ g \left(\frac{E}{E_p}\right)= \sqrt{1-\eta \left(\frac{E}{E_p}\right)^n},
\end{equation}
study of the natures of different compact objects are done in literature. Maximum possible mass acquired by a neutron star in gravity's rainbow is studied in the reference \cite{Hendi-2016}. A relation between the mass of the neutron star and Planck mass has also been established. It is shown that this relation depends on the rainbow gravity functions. Reference like \cite{Hendi-2017} show that in spite of the fact that incorporation of massive gravity modifies the FRW cosmology, it does not contradict with the idea of big bang by its own. But a careful generalization of it to the case of energy dependent space-time shows that it is possible to avoid such kind of post singularity. Authors tried to bring together all essential conditions to have a nonsingular universe. Effects of massive gravitons on different stages of evolution of universe are studied.
The proposal that the black holes can evaporate may be investigated in deeper sense by considering the rainbow gravity inspired black holes. If the specific heat of the black holes vanishes, it canbe concluded that the black hole's evaporation stops and provides the remnant mass of the concerned black hole. The Heisenberg uncertainty principle may be modified to the generalized uncertainty principle by incorporating the Planck length $l_p $ and a dimensionless positive parameter $\alpha$, named generalized uncertainty principle parameter. In the references \cite{Adler-2001, Myung-2007, Dutta-2014}, the authors have studied the black hole thermodynamics under the effect of generalized uncertainty principle. It has also been exhibited from the investigations of black holes solutions in the framework of generalized uncertainty principle that there exist of black hole remnants \cite{Adler-2001, Myung-2007, Dutta-2014, Gangopadhyay-2015}. In last few years, there have been  a lot of works done on rainbow gravity inspired black holes \cite{Ling-2007, Li-2009, Feng-2017}.

When we consider the speed of the considered object to be same of the order the speed of light we are shifted from classical mechanics to special relativity. Special relativity does not consider the order of dimensions of the considered object. But when a photon like thing is actually moving at a speed nearly equal to that of light, due to their frequency, they will observe differences in the back ground geometry. This rainbow of gravity again will depict the nature of black hole solutions in such gravity theories. Authors of the reference \cite{Panahiyan-2019} have found that the black hole solution's existence is bounded by an upper limit over rainbow functions and magnetic charge, whereas, it was limited from below by the geometrical mass. Black holes' phase transitions are found to be of first order.

Studies of thermodynamic natures and properties of different thermodynamic parameters are pursued in previous literatures.
It has been found that rainbow induced BTZ black hole's thermodynamics is deformed by rainbow functions. Existence of a remnant at the last stage of evaporation is indicated in the reference \cite{Alsaleh-2017}. Thermodynamic properties of Lovelock gravity's rainbow are studied in the reference \cite{Hendib-2017}. In this some references a critical horizon radius is found. If the original black hole's horizon radius is grater than the critical radius, it preferably becomes stable. Effects on thermal fluctions on dilatonic black holes in gravity's rainbow is studied in the reference \cite{Hendic-2017}.
But an important study is still not done. Heisenberg uncertainty principle was basically proposed to properties micro states. The generalized versions of it are efficient enough to study different macro objects. Insertion of generalized uncertainty principle may change the mode of thermodynamic studies a bit. This may enrich the stability analysis as well.
Study of properties of different phenomena regarding gravity's rainbow in the modified gravity theories rather than Einstein's general relativity has also been done. $f(R)$ gravity's rainbow in the presence of confirmally invariant Maxwell source is obtained in the reference \cite{Hendib-2016}. The geometric properties are studied and compared with the non-rainbow counter part.

This paper is organized as follows: in section $2$, we will analyze the thermodynamic properties of Reissner-Nordstr$\ddot{o}$m black hole in higher dimensions $d$ under the effects of rainbow gravity. In section $3$, we will study the same introducing the generalized uncertainty principle. In the next section, we will compute the phase transitions and examine the thermal stability of the said black hole. Finally, we will conclude in section $5$. Throughout this paper we use the Planck units, i.e.,  $G = \hbar = \kappa_B = 1$ and metric signature $(-++.....+)$.    

\section{Thermodynamics of Rainbow Gravity Inspired  Reissner-Nordstr$\ddot{o}$m Black Holes}

The Reissner-Nordstr$\ddot{o}$m black holes under the effects of rainbow gravity is given by the metric:
\begin{equation}\label{ah9.equn1}
ds^2=-\frac{1}{f^2\left(\frac{E}{E_p}\right)} \left[1-\frac{m}{r^{d-3}}+\frac{q^2}{r^{2(d-3)}}\right] dt^2+ \frac{1}{g^2\left(\frac{E}{E_p}\right)} \left[1-\frac{m}{r^{d-3}}+\frac{q^2}{r^{2(d-3)}}\right]^{-1} dr^2 + \frac{r^2}{g^2\left(\frac{E}{E_p}\right)} d\Omega_{d-2}^2,
\end{equation} 
where the constants $m$ and $q$ are respectively related to the ADM mass $M$ and the charge $Q$ of the black hole \cite{Chamblin-1999} given as:
\begin{equation}\label{ah9.equn2}
m=\frac{16\pi M}{(d-2)B_{d-2}} ~~~~~and~~~~~q=\frac{8\pi Q}{B_{d-2}\sqrt{2(d-3)(d-2)}}
\end{equation}
and $B_{d-2}$ being the volume of the $(d-2)-$sphere expressed as
\begin{equation}\label{ah9.equn3}
 B_{d-2}=\frac{2\pi^\frac{d-1}{2}}{\Gamma\left(\frac{d-1}{2}\right)}~.
\end{equation}
From equation (\ref{ah9.equn1}) it may be apparently seem that the rainbow functions can be adjustable with the spatial $r$ and temporal $t$ coordinates as these functions are not depending on coordinates explicitly. But these functions are dependent on the energy carried by the concerned particle which again may be treated as the combination of displacement and velocity (or time derivative of displacement). This is why even if we absorb the rainbow functions in the coordinates, this will not give a metric exactly equivalent to the general relativistic counter part of it. This is why the existence of these functions (when $\neq 1$) may give rise to drastic changes to the thermodynamic studies.
The event horizon of the black hole may be computed from the lapse function as
\begin{equation}\label{ah9.equn4}
r_h=\left[ \frac{m}{2}+ \frac{\sqrt{m^2-4q^2}}{2}\right]^{\frac{1}{d-3}}.
\end{equation}

The surface gravity $ \kappa $ of the black holes related to the temperature $ T_h $ of black hole at the event horizon as per the equation \cite{Bekenstein-1974, Hawking-1975, Hawking-1976}:
\begin{equation}\label{ah9.equn5}
T_h=\frac{\kappa}{2\pi}
\end{equation}
and the surface gravity $ \kappa $ is defined by
\begin{equation}\label{ah9.equn6}
\kappa=\lim_{r\longrightarrow r_h} \frac{1}{2}\left[\sqrt{-{g^{rr} g_{tt}(g_{tt,r})^2}}\right],
\end{equation}
which gives

\begin{equation}\label{ah9.equn7}
\kappa= \frac{f\left(\frac{E}{E_p}\right)}{g\left(\frac{E}{E_p}\right)}\left( \frac{d-3}{2}\right) \left[\frac{m}{r_h^{d-2}} -\frac{2 q^2}{r_h^{2d-5}}\right]
\end{equation}
for the particular kind of black holes we are discussing about in this article.\\
The Hawking temperature, therefore is given by

$$T_h=\frac{d-3}{4 \pi}\sqrt{1- \frac{\eta}{E_p^n} \frac{1}{r_h^n}} \left(\frac{m}{r_h^{d-2}} -\frac{2 q^2}{r_h^{2d-5}}\right)$$

\begin{equation}\label{ah9.equn8}
=\frac{d-3}{4 \pi}\sqrt{1- \frac{\eta}{E_p^n} \frac{1}{\left[ \frac{m}{2}+ \frac{\sqrt{m^2-4q^2}}{2}\right]^{\frac{n}{d-3}}}} \left(\frac{m}{\left[ \frac{m}{2}+ \frac{\sqrt{m^2-4q^2}}{2}\right]^{\frac{{d-2}}{d-3}}} -\frac{2 q^2}{\left[ \frac{m}{2}+ \frac{\sqrt{m^2-4q^2}}{2}\right]^{\frac{2d-5}{d-3}}}\right).
\end{equation}
According to the heuristic arguments in the references \cite{Adler-2001, Cavaglia-2004}, even in rainbow gravity, Heisenberg uncertainty principle holds, i.e., $\Delta x \Delta p \geq 1$ still holds. This is why we are able to translate Heisenberg uncertainty principle into a lower bound on the energy. Mathematically, $E \geq \frac{1}{\Delta x}$, where $E$ is the particles energy emitted as the Hawking radiation. One can obtain now the lower bound of the energy as \cite{Alib-2014}: $E \geq 1 \approx \frac{1}{r_h}$. This is why, to obtain the above relation we have put $E=\frac{1}{r_h}$.

\begin{figure}[h!]
\begin{center}
~~~~~~~~~~~~~~~~~~Fig.-1a~~~~~~~~~~~~~~~~~~~~~~~~~~~~~~~~~~~~~~Fig.-1b~~~~~~~~~~~~~~~~~~~~~~\\
\includegraphics[scale=.7]{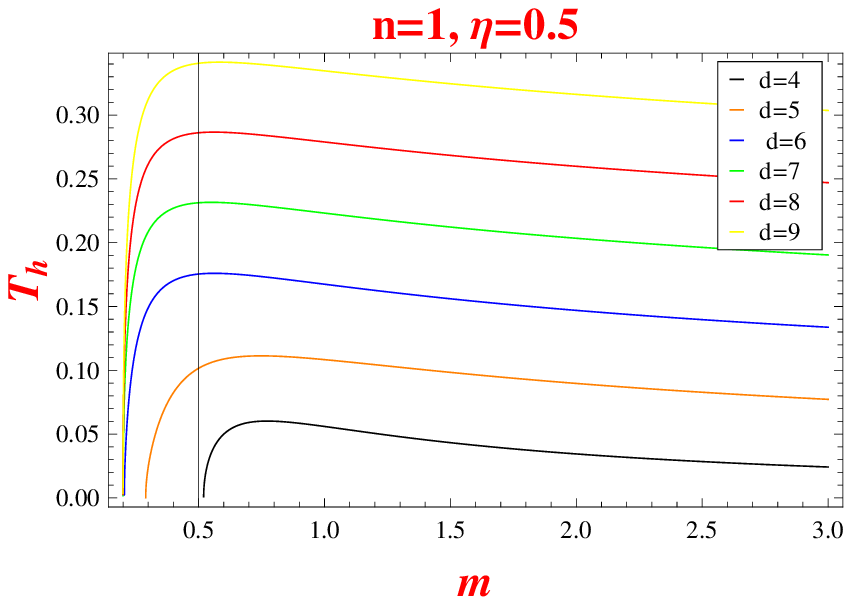}
\includegraphics[scale=.7]{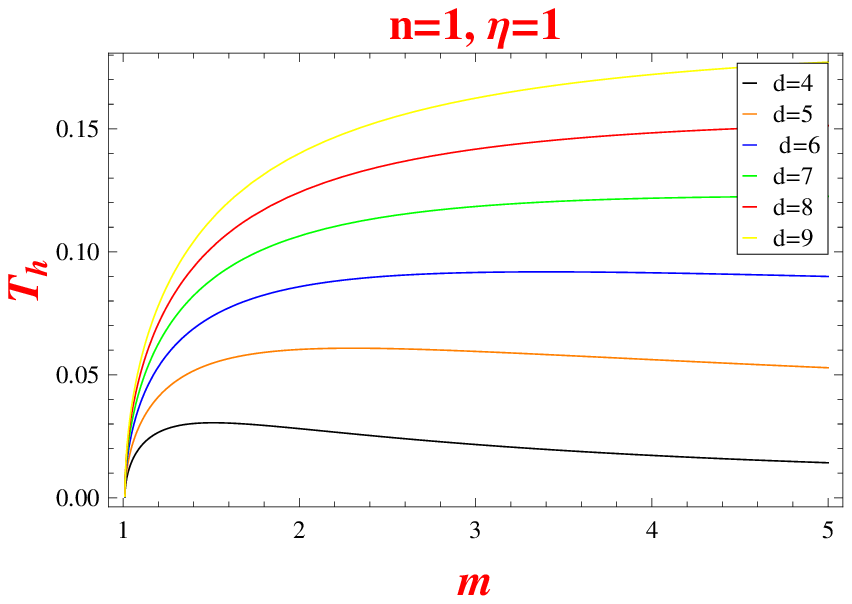}

~~~~~~~~~~~~~~~~~~Fig.-2a~~~~~~~~~~~~~~~~~~~~~~~~~~~~~~~~~~~~~~Fig.-2b~~~~~~~~~~~~~~~~~~~~~~~~~\\
\includegraphics[scale=.7]{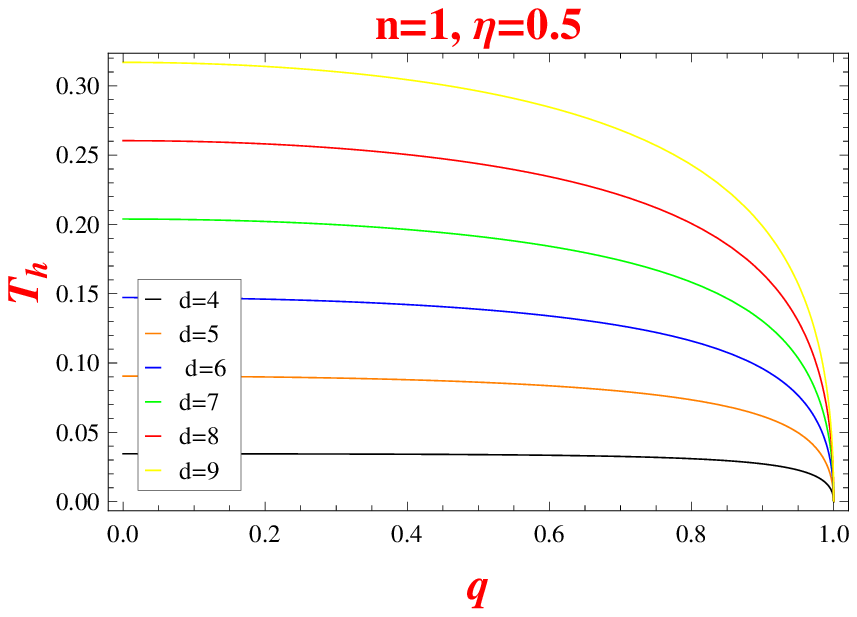}
\includegraphics[scale=.7]{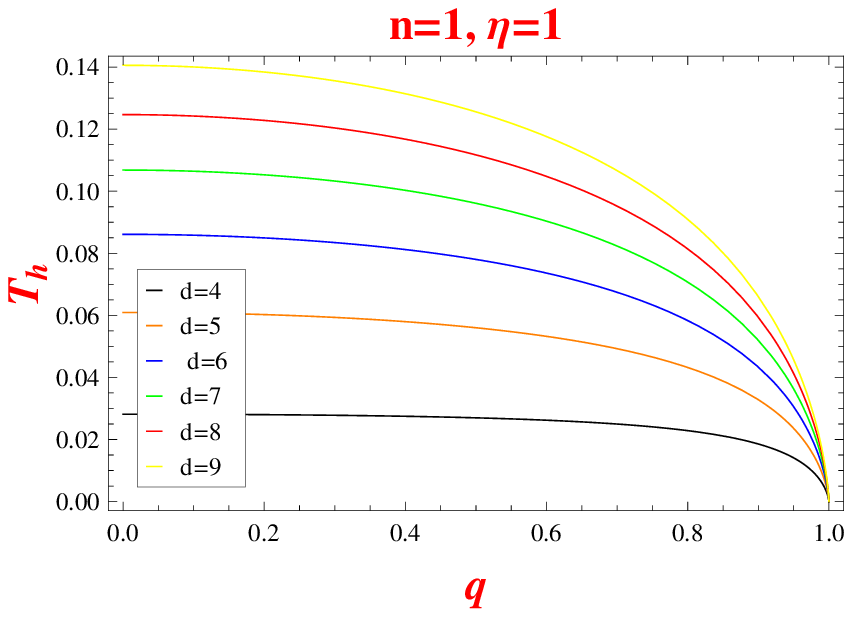}

Fig.-1a and 1b represent the variations of $ T_h $ with respect to $ m $ in rainbow gravity with $\eta= 0.5$ and $1$ respectively for different dimensions $d$, $\eta$ is taken to be $1$ and $q=0.1$ .\\
Fig.-2a and 2b represent the variations of $ T_h $ with respect to $ q $ in rainbow gravity with $\eta= 0.5$ and $1$ respectively for different dimensions $d$, $\eta$ is taken to be $1$ and $m=2$ .
\end{center} 
\end{figure}
Figures 1a and 1b depict the variations of horizon temperature $ T_h $ with respect to $m$ for different values of rainbow gravity parameter $\eta$ when $n$ is taken to be equal to 1. The vertical black line appears in the graph 1a is $m=0.5$ line. This has been drawn to locate the values $m_{crit}$ for which $T_h$ reaches to a local minima. We can easily point out that $m_{crit}(d_1)>m_{crit}(d_2)$ if $d_1<d_2$. For low $\eta$, we observe that the temperature rises to a local maxima and then reduces and becomes assymptotic to $m$ axis. If we increase the dimension $d$, the maxima increases and it occurs at a lower value of $m$. Increment in $\eta$ is followed by an ``only increasing curve", when the dimension $d$ is high. For low dimension $d$, however, the nature is almost like low $\eta$ case. If we look at the equation (\ref{ah9.equn8}), we observe that if the dimension $d$ is less, the quantity $\frac{d-2}{d-3}$ has a value grater than unity and as a result the increment of $T_h$ stops if $m$ increases abruptly. On the other hand if $d$ is high, $\frac{d-2}{d-3}\rightarrow 1$ and $T_h$ is ever increasing. We can clarify this result physically as well. If we increase the dimension $d$, it is likely to have a more permeable event horizon, i.e., the black holes will become more unstable and naked singularity may arise \cite{Rudraa 2011, Debnath 2012, Rudrab 2012, Rudrac 2014}. Now instability arises due to high temperature which tries to radiate the black hole into the corresponding hot AdS space. So high $d$ naturally gives rise to high temperature $T_h$. On the other hand, increment in $\eta$ shifts us from the special relativity and turns us to a distorted gravity which arises mainly due to the energy differences of the photons. Change in the value of $\eta$ is brought through the change in $g\left(\frac{E}{E_p}\right)$. Again change in $g\left(\frac{E}{E_p}\right)$ is directly related to the change of the lapse function $g_{rr}$ in equation (\ref{ah9.equn1}). So if $\eta$ is less, $g\left(\frac{E}{E_p}\right)$ stays near to the special relativistic region and hence we can stay near to Einstein gravity. This is why low dimension $d$ and lower value of $\eta$ show same thermodynamic natures. If we make $m$ constant and vary the charge $q$ (Figures 2a and 2b), it will be observed that for low $\eta$ the  horizon temperature $ T_h $ reduces for increasing $q$ and whatever be the value of $d$, after a while all the curves converge to a same point. The rate of decrease increases if we increase the value of $\eta$. 

Physically, this says low $\eta$ and low $d$ together support phase transition. But if $\eta$ and $d$ both are high, no phase transition is obtained and the larger black hole is made stable. Black hole with high charge, however, does not exist. There exists a particular upper boundary of charge beyond which, whatever be the value of $d$, we will not find positive horizon temperature $T_h$, i.e., a physical black hole.

Since the temperature must be a real quantity, we must have
  
\begin{equation}\label{ah9.equn9}
1- \frac{\eta}{E_p^n} \frac{1}{r_h^n}\geq 0
\end{equation}
and this term provides the critical mass below which the temperature becomes complex. The critical mass of the Reissner-Nordstr$\ddot{o}$m black hole in higher dimensions under the effects of rainbow gravity is therefore expressed as
\begin{equation}\label{ah9.equan10}
m_{critical}=q^2 \left(\frac{E_p^n}{\eta }\right)^{\frac{d-3}{n}}+\left(\frac{E_p^n}{\eta }\right)^{\frac{3-d}{n}}.
\end{equation}
The heat capacity of this black hole is given as
\begin{equation}\label{ah9.equn11}
C_h=\frac{dm}{dT_h}
\end{equation}
and we obtain the equation (\ref{ah9.equn31}) of Appendix for the heat capacity.

Putting $C_h=0$, in equation (\ref{ah9.equn31}) one can obtain the remnant mass of this black hole as
\begin{equation}\label{ah9.equn12}
 m_{remnant}=q^2 \left(\frac{E_p^n}{\eta }\right)^{\frac{d-3}{n}}+\left(\frac{E_p^n}{\eta }\right)^{\frac{3-d}{n}}
\end{equation}
and this is found to be equal to the critical mass.  

\begin{figure}[h!]
\begin{center}
~~~~~~~~~~~~~~~~~~Fig.-3a~~~~~~~~~~~~~~~~~~~~~~~~~~~~~~~~~~~~~~Fig.-3b~~~~~~~~~~~~~~~~~~~~~~\\
\includegraphics[scale=.7]{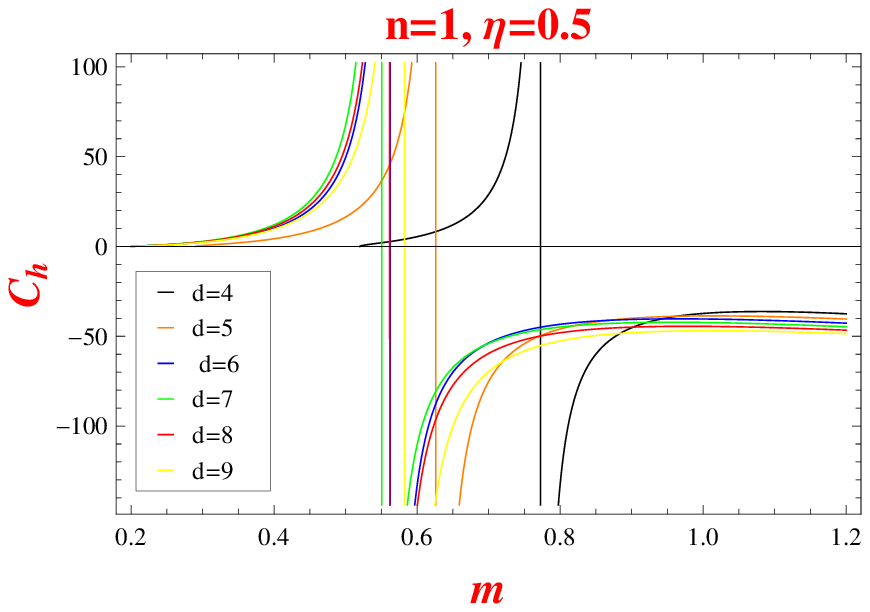}
\includegraphics[scale=.7]{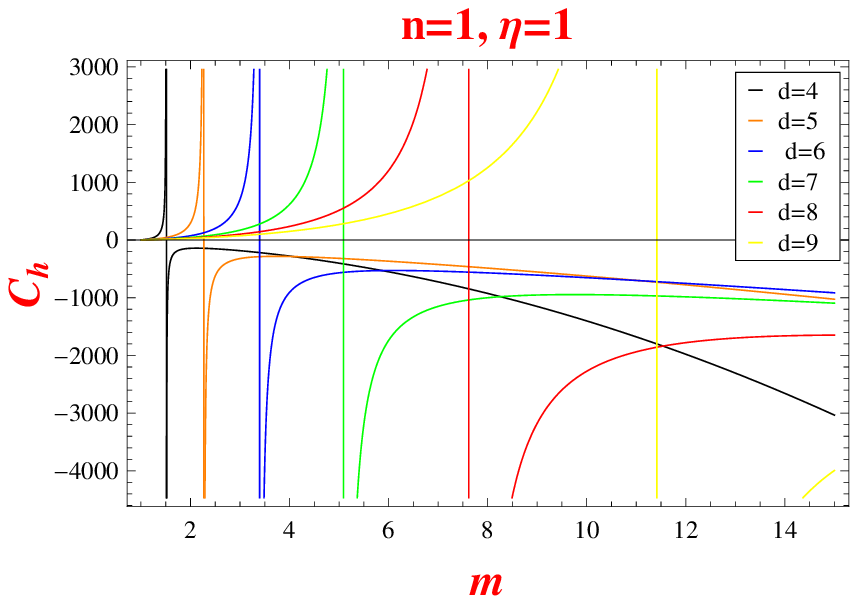}

~~~~~~~~~~~~~~~~~~Fig.-4a~~~~~~~~~~~~~~~~~~~~~~~~~~~~~~~~~~~~~~Fig.-4b~~~~~~~~~~~~~~~~~~~~~~~~~\\
\includegraphics[scale=.7]{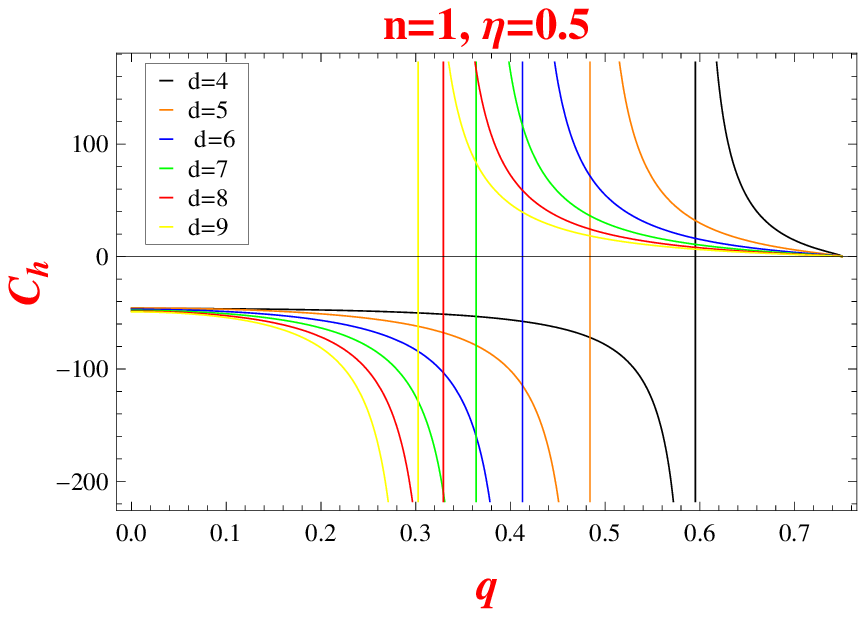}
\includegraphics[scale=.7]{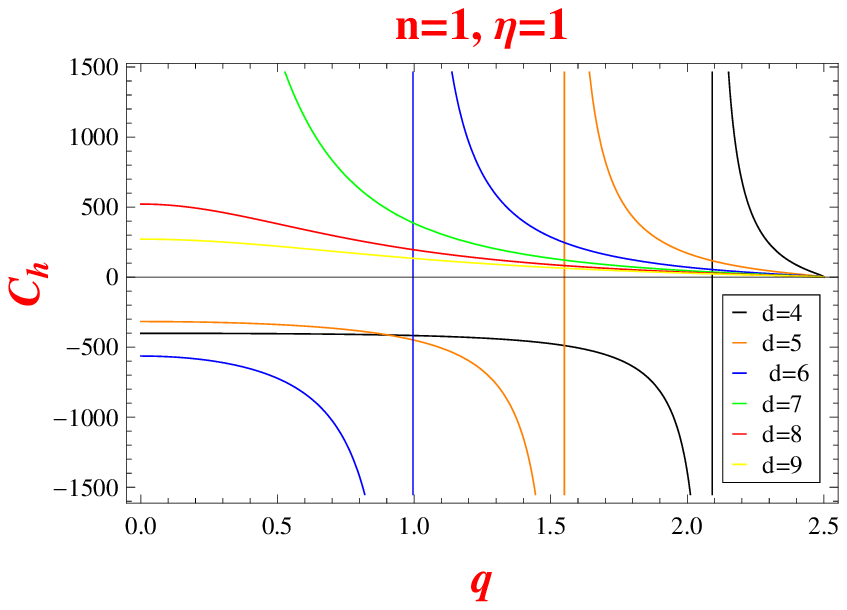}

Fig.-3a and 3b represent the variations of $ C_h $ with respect to $ m $ in rainbow gravity with $\eta= 0.5$ and $1$ respectively for different dimensions $d$, $\eta$ is taken to be $1$ and $q=0.1$.\\
Fig.-4a and 4b represent the variations of $ C_h $ with respect to $ q $ in rainbow gravity with $\eta= 0.5$ and $1$ respectively for different dimensions $d$, $\eta$ and $r_h=1$ are taken to be $1$ .
\end{center} 
\end{figure}
Figures 3a and 3b show that the specific heat at horizon $C_h$ says the black hole goes through a phase transition which transits a stable black hole to its unstable counter part. For high $\eta$, the point of phase transition occurs at high $m$ if $d$ is increased.

Increment in charge (Figures 4a and 4b), however, converts an unstable black hole into a stable one.

Entropy is always increasing supporting generalized second law of thermodynamics (Figures 5a to 6b).

The entropy of this black hole in higher dimensions under rainbow gravity effects is now computed as:

\begin{equation}\label{ah9.equn13}
S_h= \int \frac{dm}{\frac{d-3}{4 \pi}\sqrt{1- \frac{\eta}{E_p^n} \frac{1}{r_h^n}} \left(\frac{m}{r_h^{d-2}} -\frac{2 q^2}{r_h^{2d-5}}\right)}
\end{equation}
as we find the equation (\ref{ah9.equn32}) of Appendix for the entropy of the black hole.

\begin{figure}[h!]
\begin{center}
~~~~~~~~~~~~~~~~~~Fig.-5a~~~~~~~~~~~~~~~~~~~~~~~~~~~~~~~~~~~~~~Fig.-5b~~~~~~~~~~~~~~~~~~~~~~\\
\includegraphics[scale=.7]{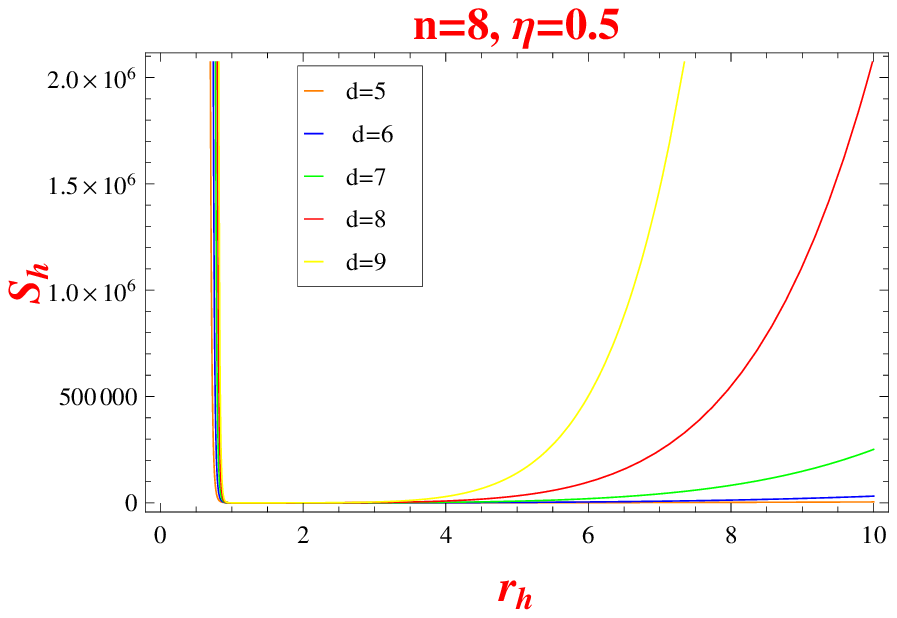}
\includegraphics[scale=.7]{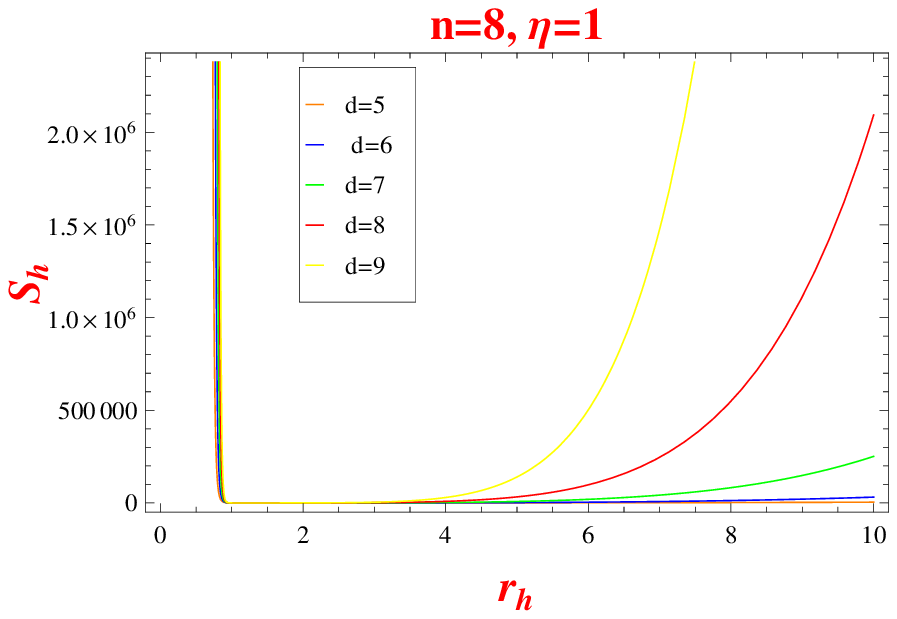}

~~~~~~~~~~~~~~~~~~Fig.-6a~~~~~~~~~~~~~~~~~~~~~~~~~~~~~~~~~~~~~~Fig.-6b~~~~~~~~~~~~~~~~~~~~~~~~~\\
\includegraphics[scale=.7]{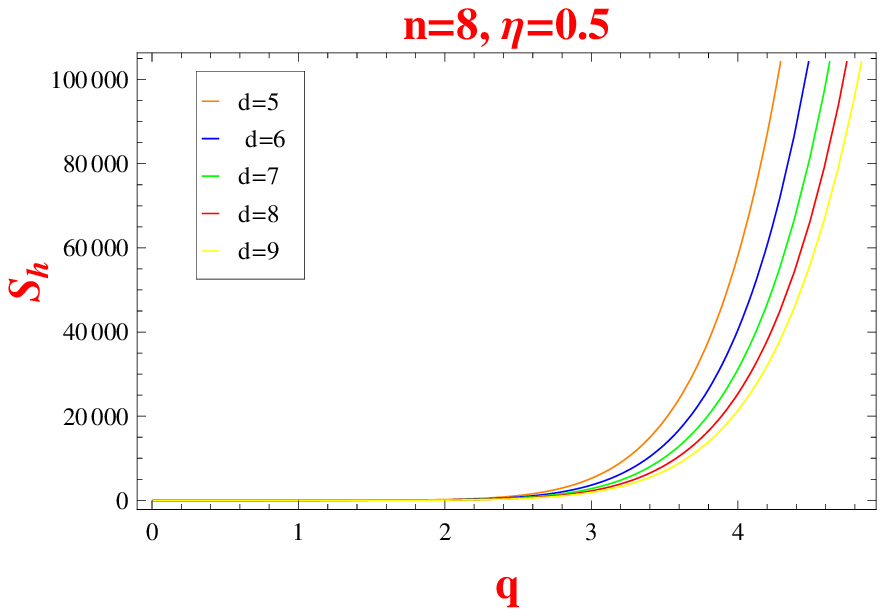}
\includegraphics[scale=.7]{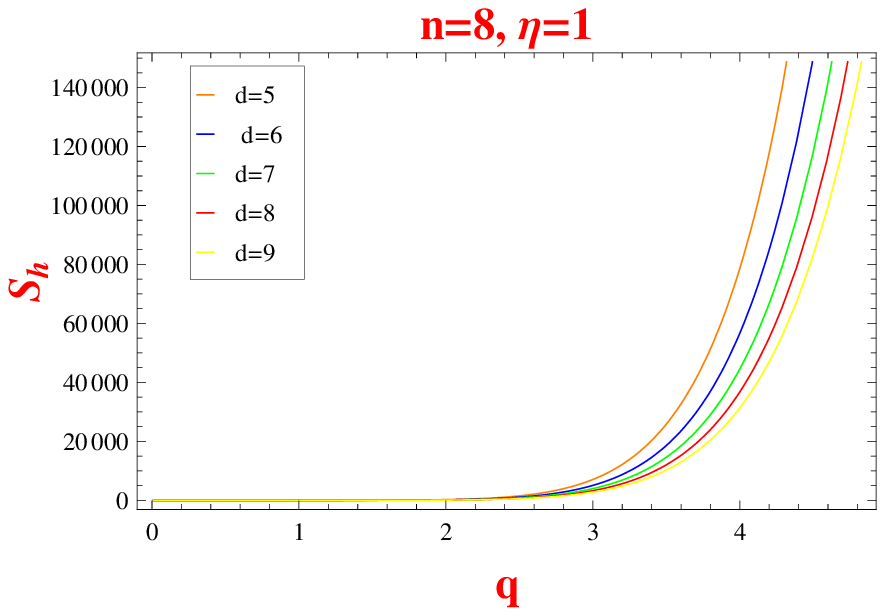}

Fig.-5a and 5b represent the variations of $ S_h $ with respect to $ r_h $ in rainbow gravity with $\eta= 0.5$ and $1$ respectively for different dimensions $d$, $\eta$ is taken to be $1$ and $q=0.1$.\\
Fig.-6a and 6b represent the variations of $ S_h $ with respect to $ q $ in rainbow gravity with $\eta= 0.5$ and $1$ respectively for different dimensions $d$, $\eta$ is taken to be $1$ and $m=2$ and and $r_h=1$.
\end{center} 
\end{figure}
\section{Thermodynamics of Reissner-Nordstr$\ddot{o}$m Black Holes Under the Effects of Rainbow Gravity with General Uncertainty Principle}

The generalized form of Heisenberg's uncertainty principle in the microphysics regime may be stated as \cite{Zhao-2003, Sun-2004, Kim-2007, Yoon-2007, Anacieto-2014, Nourcer-2007, Anacieto-2015, Anacieto-2016}:
\begin{equation}\label{ah9.equn14}
\triangle x \triangle p\geq \left(1-\alpha l_p \triangle p+ \alpha^2 l_p^2 \triangle p^2\right),
\end{equation}
where $l_p= \sqrt{\frac{\hbar G}{c^3}}$ is the Planck's length of the order of $10^{-35}m$ and $\alpha$ is a dimensionless positive parameter. To obtain the quantum corrected thermodynamic parameters, we apply the quadratic form of GUP and this is expressed as

\begin{equation}\label{ah9.equn15}
\triangle x \triangle p\geq \left(1+\alpha^2 l_p^2 \triangle p^2\right)
\end{equation}
 which may also be rewritten as:
\begin{equation}\label{ah9.equn16}
\triangle p \geq \frac{\triangle x}{2\alpha^2 l_p^2}\left(1-\sqrt{1-\frac{4\alpha^2 l_p^2}{\triangle x^2}} \right).
\end{equation}

Taking $\frac{l_p}{\triangle x}\ll 1$ and using Taylor series expansion,  the equation (\ref{ah9.equn16}) takes the form as: 
\begin{equation}\label{ah9.equn17}
\triangle p \geq \frac{1}{\triangle x}\left(1+\frac{2\alpha^2 l_p^2}{\triangle x^2}+... \right).
\end{equation}
Applying the saturated form of the uncertainty principle $\triangle x \triangle p\geq 1$, we have $E\triangle x\geq 1$ \cite{Haldar 2018, Ali 2015}. Then the equation (\ref{ah9.equn17}) can be written as:
\begin{equation}\label{ah9.equn18}
E_G\geq E \left(1+\frac{2\alpha^2 l_p^2}{\triangle x^2}+... \right).
\end{equation} 
Therefore, for a particle with corrected energy $E_G$ crossing the event horizon the tunneling probability  will be,
\begin{equation}\label{ah9.equn19}
\Gamma \simeq exp [-2 Im {\cal I_G}]=exp\left\{-\frac{4\pi E_G}{f^{'}(r_h)}\right\}.
\end{equation}
Comparing the equation (\ref{ah9.equn19}) with the Boltzmann factor $exp \{-\frac{E}{T}\}$, one can obtain the quantum-corrected Hawking temperature as \cite{Haldar 2018}:
\begin{equation}\label{ah9.equn20}
T_G=T_h\left(1+\frac{2\alpha^2 l_p^2}{\triangle x^2}+... \right)^{-1},
\end{equation}
where $T_h$ is given as  equation (\ref{ah9.equn8}) and we consider that the position uncertainty  for events near the event horizon is given by $\triangle x\simeq 2r_h$ \cite{Medved-2004}. Thus the quantum-corrected temperature of the black hole is given by
\begin{equation}\label{ah9.equn21}
T_G=T_h\left(1+\frac{\alpha^2 l_p^2}{2r_h^2}+... \right)^{-1}
\end{equation}
which gives the equation (\ref{ah9.equn33}) of Appendix for the  quantum-corrected temperature of the black hole.
\begin{figure}[h!]
\begin{center}
~~~~~~~~~Fig.-7a~~~~~~~~~~~~~~~~~~~~~~~~~Fig.-7b~~~~~~~~~~~~~~~~~~~~~~Fig.-7c~~~~~~~~~~~~~~~~~~\\
\includegraphics[scale=.5]{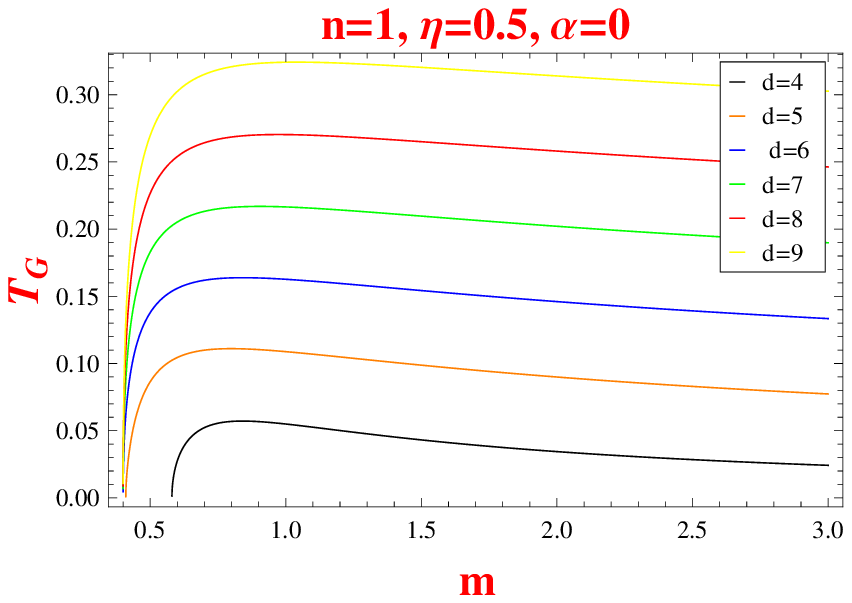}
\includegraphics[scale=.5]{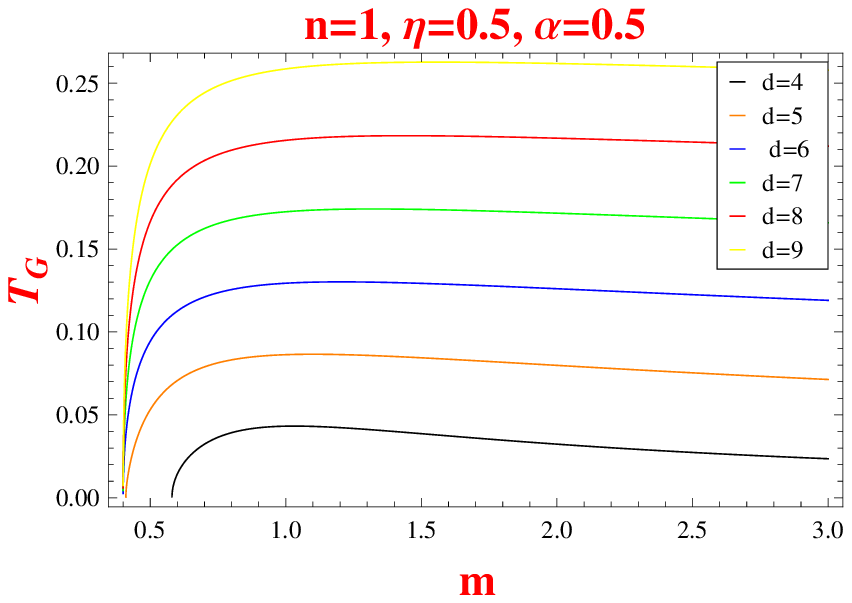}
\includegraphics[scale=.5]{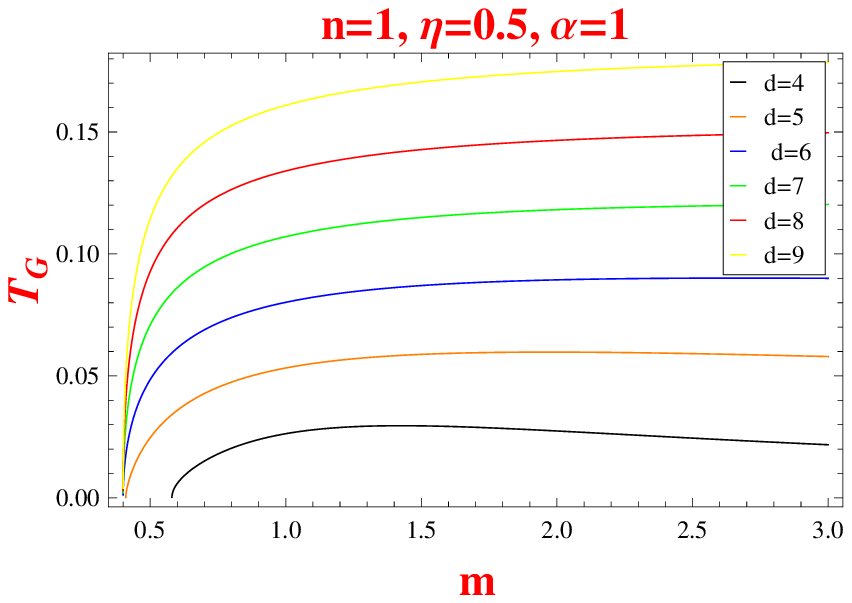}

~~~~~~~~Fig.-8a~~~~~~~~~~~~~~~~~~~~~~~~~Fig.-8b~~~~~~~~~~~~~~~~~~~~~~~Fig.-8c~~~~~~~~~~~~~~~~~\\
\includegraphics[scale=.5]{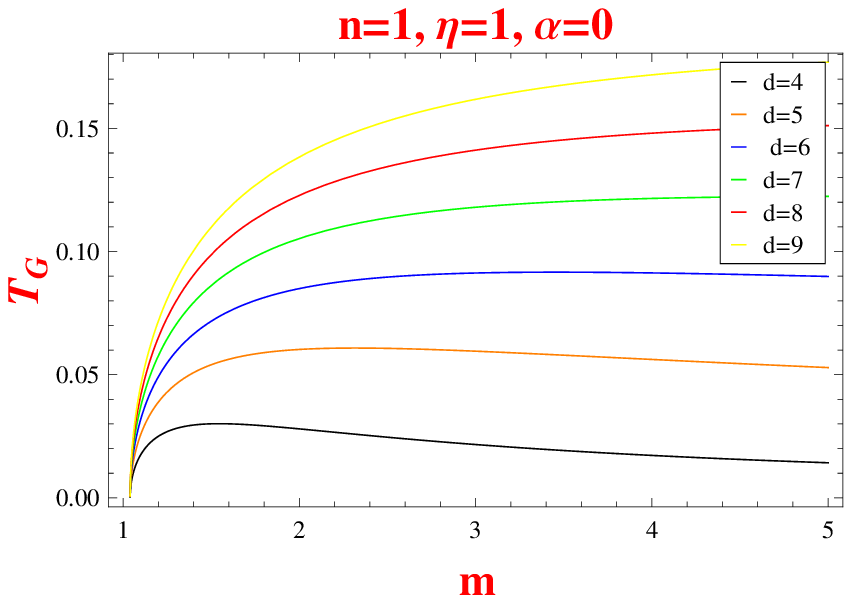}
\includegraphics[scale=.5]{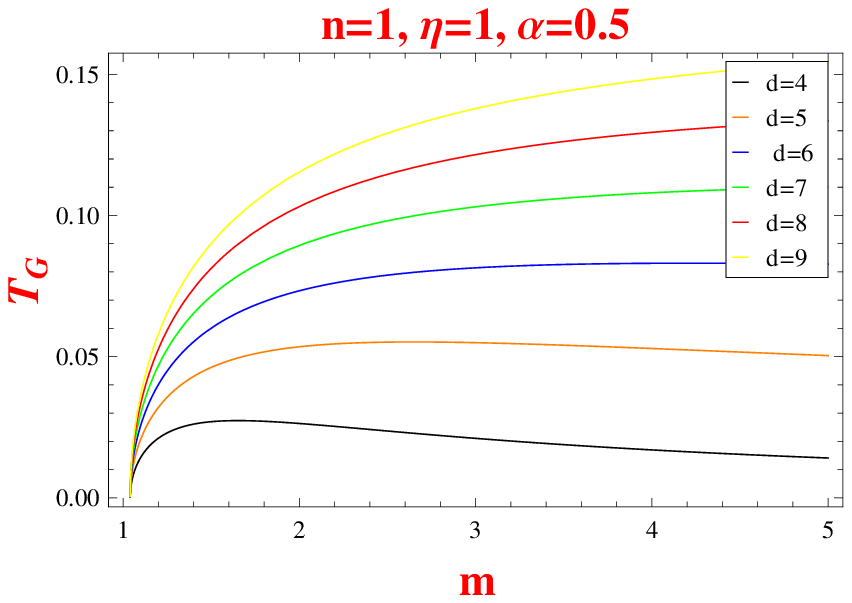}
\includegraphics[scale=.5]{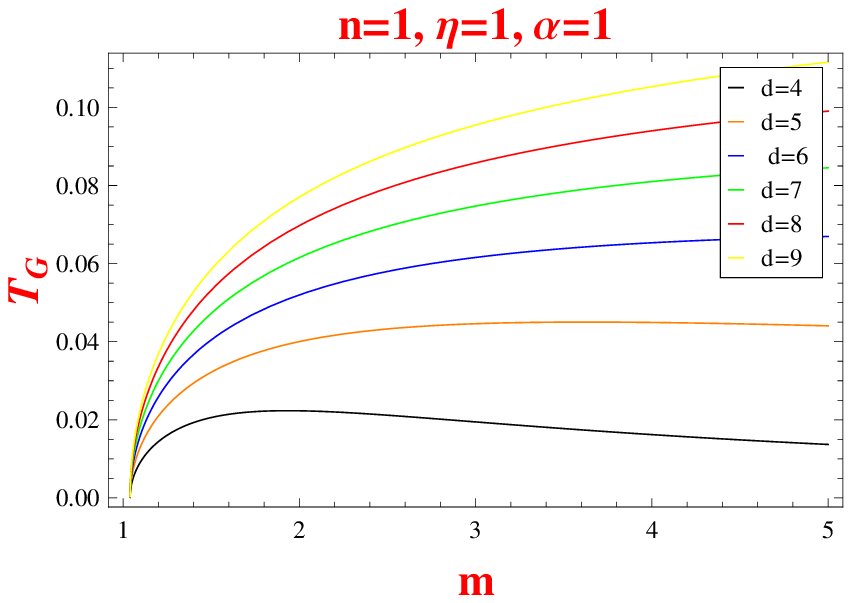}

Fig.-7a-7c represent the variations of $ T_G $ with respect to $ m $ in rainbow gravity with $\eta= 0.5$ for different dimensions $d$ and $\alpha=0, 0.5$ and $1$ and $q=0.1$.\\
Fig.-8a-8c represent the variations of $ T_G $ with respect to $ m $ in rainbow gravity with $\eta= 1$ for different dimensions $d$ and $\alpha=0, 0.5$ and $1$ and $q=0.1$.
\end{center} 
\end{figure}

\begin{figure}[h!]
\begin{center}
~~~~~~~~~Fig.-9a~~~~~~~~~~~~~~~~~~~~~~~~~Fig.-9b~~~~~~~~~~~~~~~~~~~~~~Fig.-9c~~~~~~~~~~~~~~~~~~\\
\includegraphics[scale=.5]{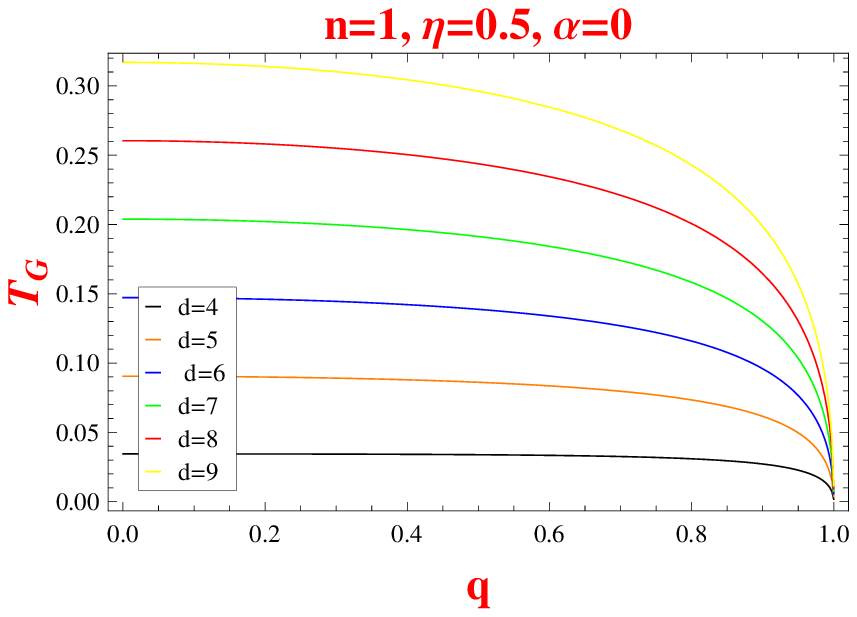}
\includegraphics[scale=.5]{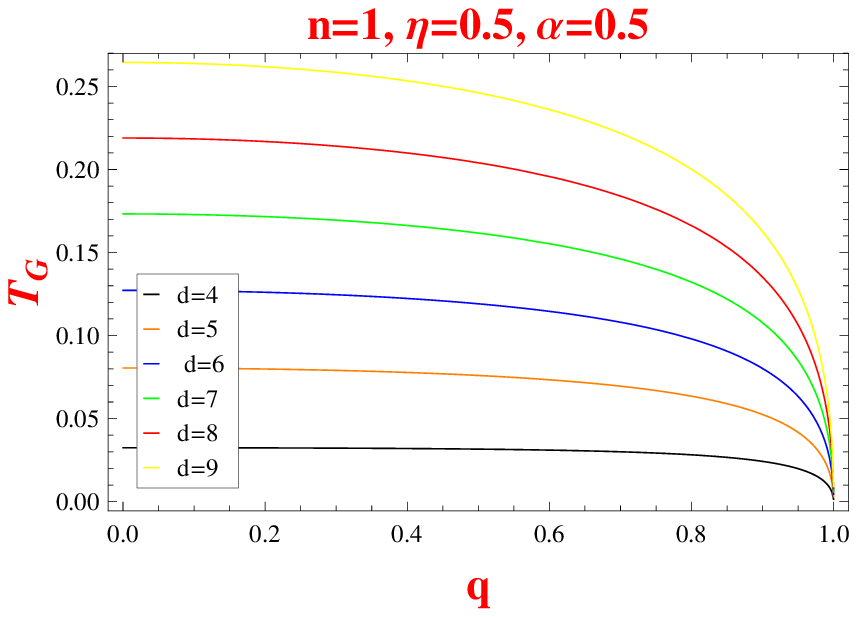}
\includegraphics[scale=.5]{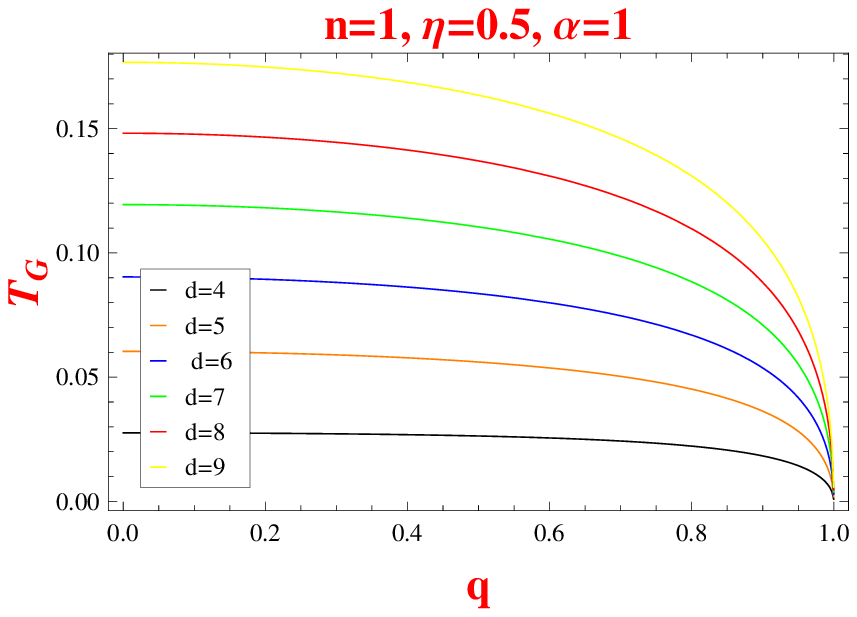}

~~~~~~~~Fig.-10a~~~~~~~~~~~~~~~~~~~~~~~~~Fig.-10b~~~~~~~~~~~~~~~~~~~~~~~Fig.-10c~~~~~~~~~~~~~~~~~\\
\includegraphics[scale=.5]{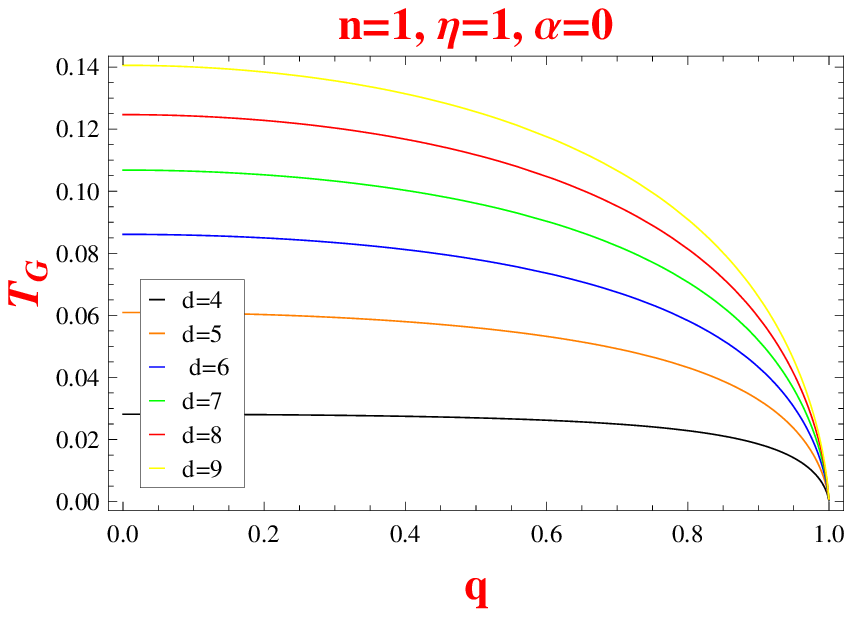}
\includegraphics[scale=.5]{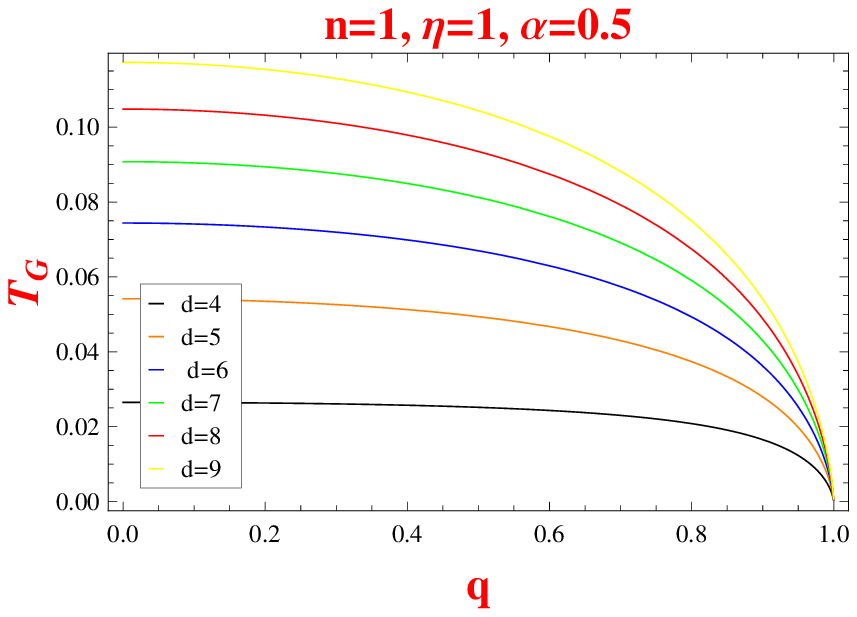}
\includegraphics[scale=.5]{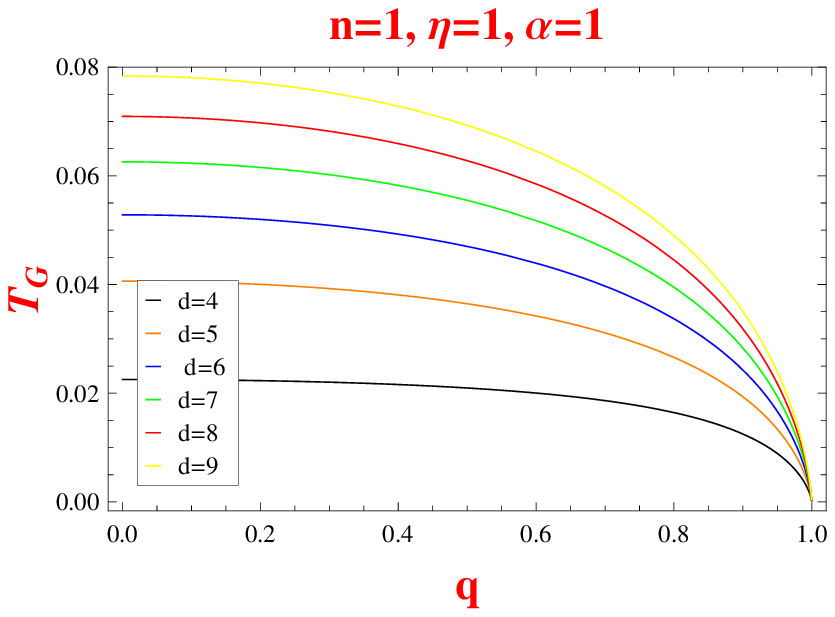}

Fig.-9a-9c represent the variations of $ T_G $ with respect to $ q $ in rainbow gravity with $\eta= 0.5$ for different dimensions $d$ and $\alpha=0, 0.5$ and $1$ and $m=2$.\\
Fig.-10a-10c represent the variations of $ T_G $ with respect to $ q $ in rainbow gravity with $\eta= 1$ for different dimensions $d$ and $\alpha=0, 0.5$ and $1$ and $m=2$.
\end{center} 
\end{figure}
and hence the quantum corrected heat capacity of the black hole is given by
\begin{equation}\label{ah9.equn22}
C_G=\frac{dm}{dT_G}
\end{equation}
which takes the form of equation (\ref{ah9.equn34}) of Appendix for  the quantum corrected heat capacity of the black hole.
It is evident from the equation (\ref{ah9.equn34}) that for the limit $\alpha \rightarrow 0$, $C_G$ reduces to $C_h$ as shown in equation (\ref{ah9.equn11}).

\begin{figure}[h!]
\begin{center}
~~~~~~~~~Fig.-11a~~~~~~~~~~~~~~~~~~~~~~~~~Fig.-11b~~~~~~~~~~~~~~~~~~~~~~Fig.-11c~~~~~~~~~~~~~~~~~~\\
\includegraphics[scale=.5]{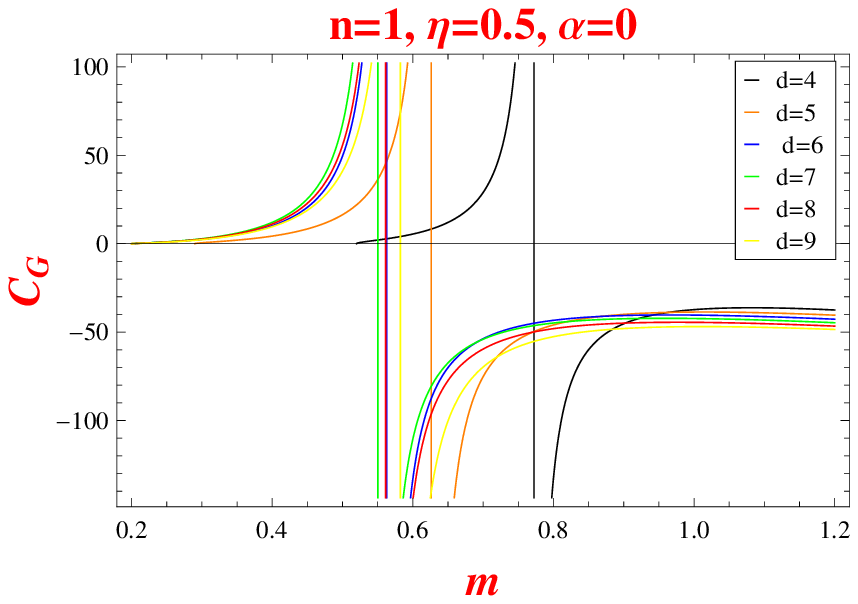}
\includegraphics[scale=.5]{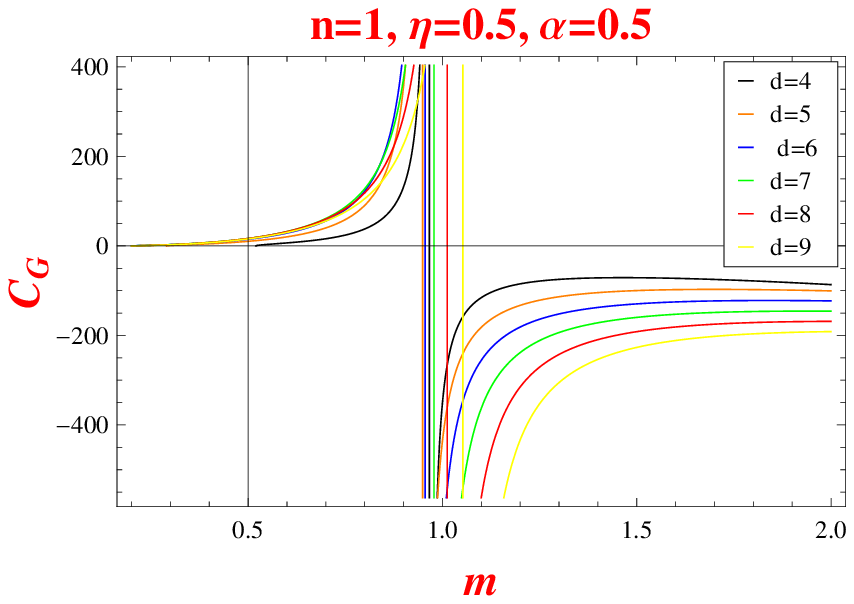}
\includegraphics[scale=.5]{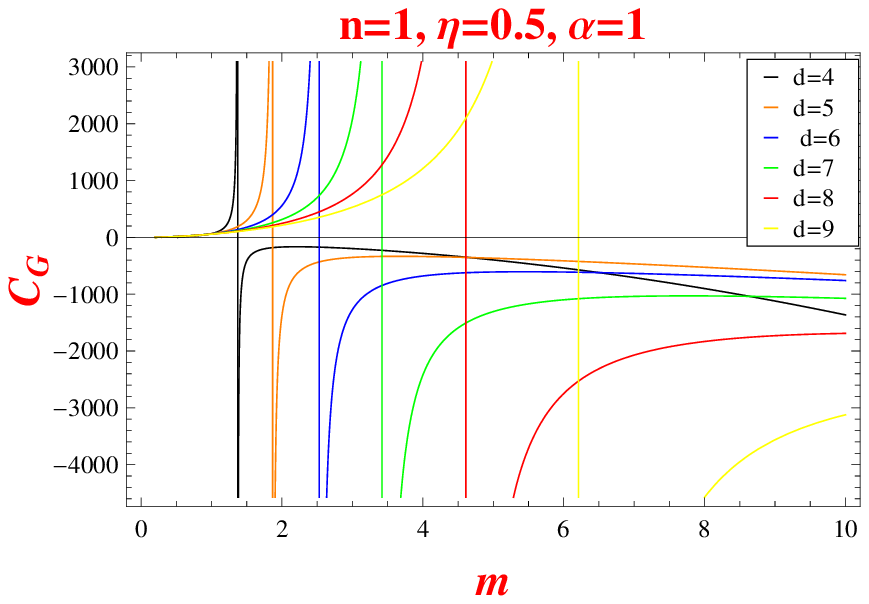}

~~~~~~~~Fig.-12a~~~~~~~~~~~~~~~~~~~~~~~~~Fig.-12b~~~~~~~~~~~~~~~~~~~~~~~Fig.-12c~~~~~~~~~~~~~~~~~\\
\includegraphics[scale=.5]{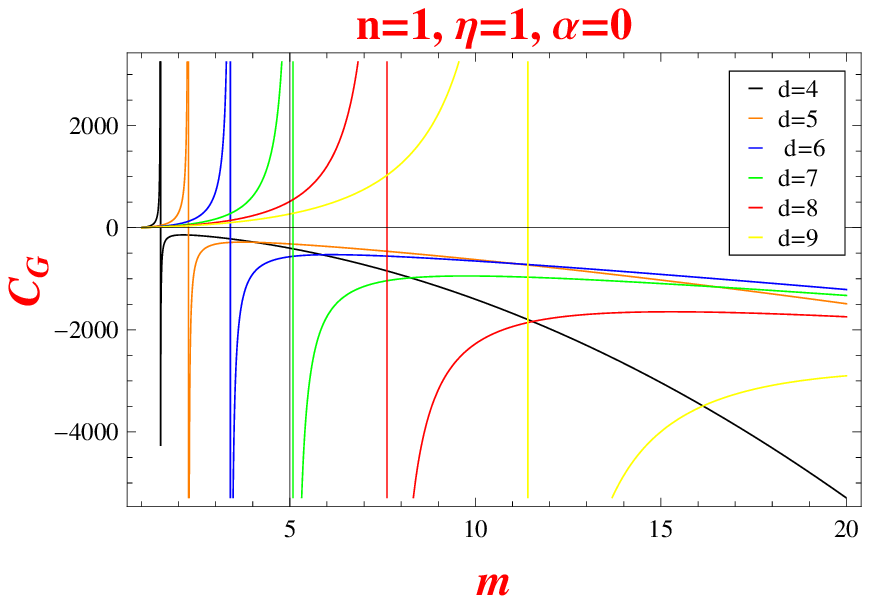}
\includegraphics[scale=.5]{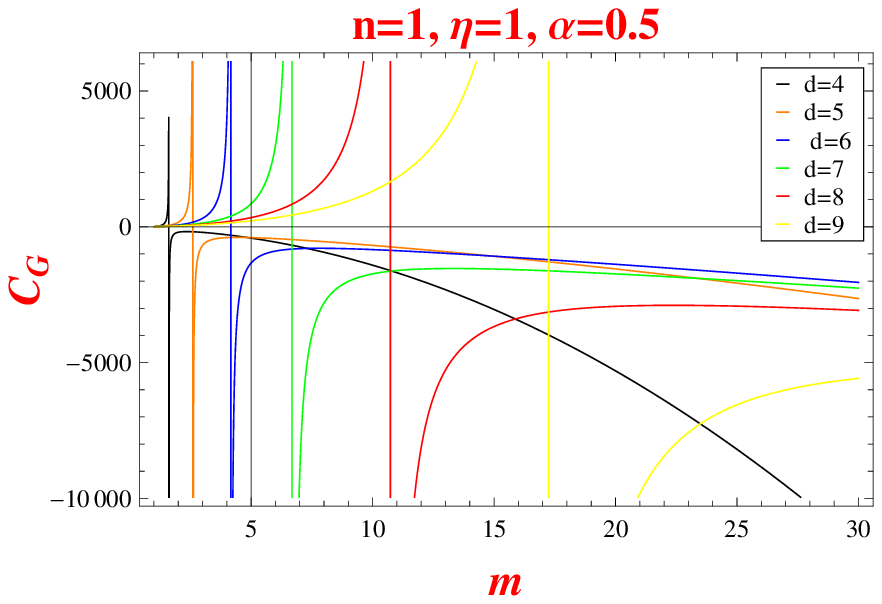}
\includegraphics[scale=.5]{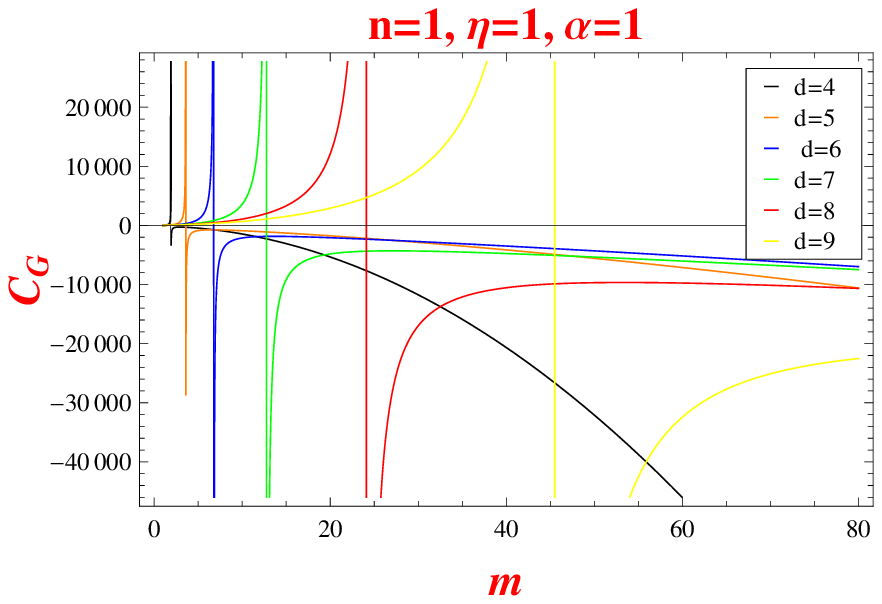}

Fig.-11a-11c represent the variations of $ T_G $ with respect to $ m $ in rainbow gravity with $\eta= 0.5$ for different dimensions $d$ and $\alpha=0, 0.5$ and $1$ and $q=0.1$.\\
Fig.-12a-12c represent the variations of $ T_G $ with respect to $ m $ in rainbow gravity with $\eta= 1$ for different dimensions $d$ and $\alpha=0, 0.5$ and $1$ and $q=0.1$.
\end{center} 
\end{figure}

\begin{figure}[h!]
\begin{center}
~~~~~~~~~Fig.-13a~~~~~~~~~~~~~~~~~~~~~~~~~Fig.-13b~~~~~~~~~~~~~~~~~~~~~~Fig.-13c~~~~~~~~~~~~~~~~~~\\
\includegraphics[scale=.5]{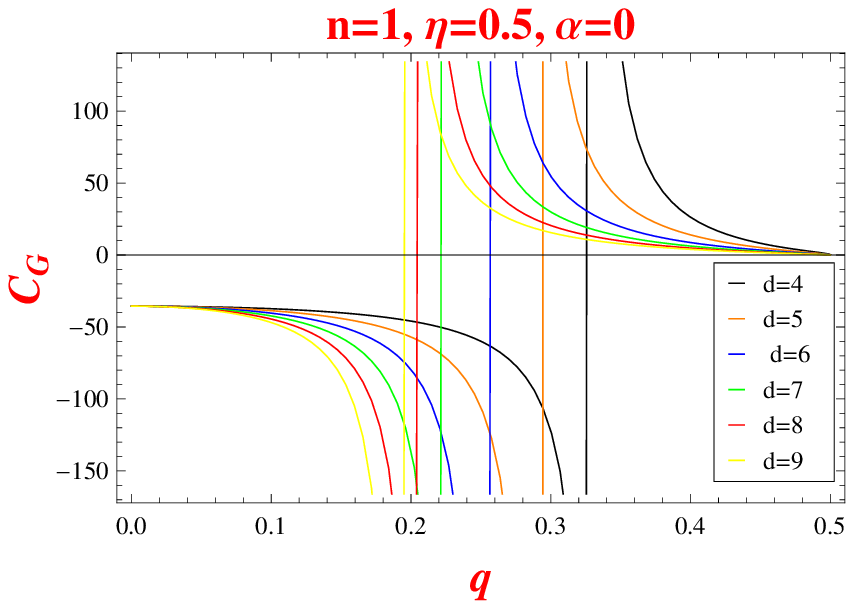}
\includegraphics[scale=.5]{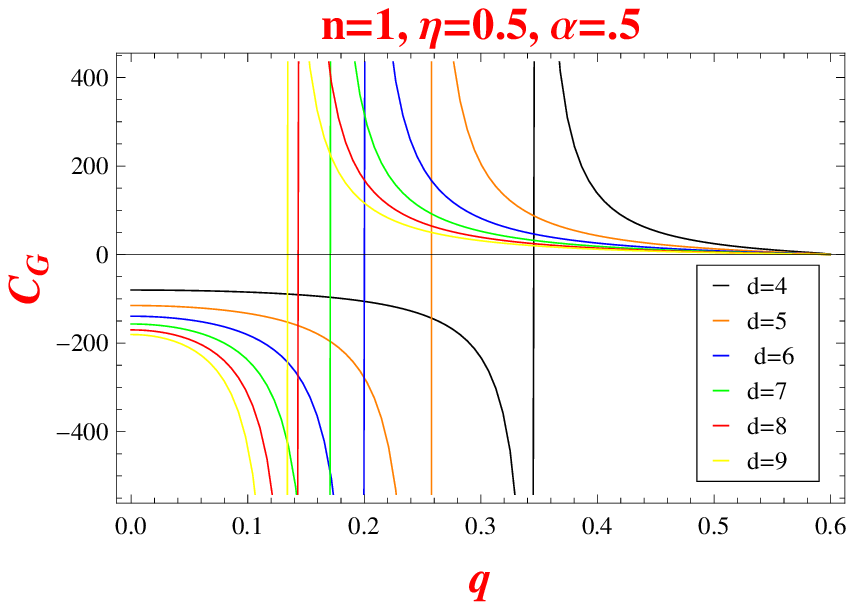}
\includegraphics[scale=.5]{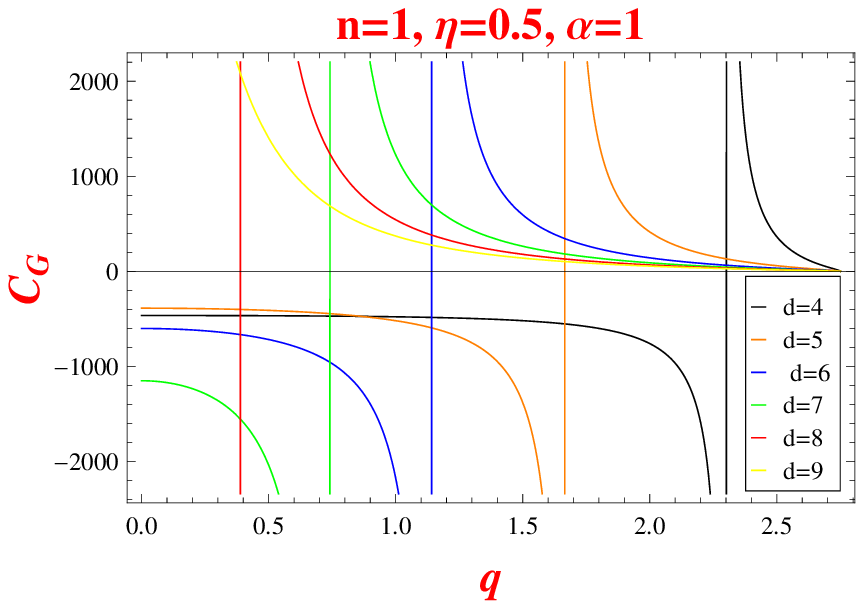}

~~~~~~~~Fig.-14a~~~~~~~~~~~~~~~~~~~~~~~~~Fig.-14b~~~~~~~~~~~~~~~~~~~~~~~Fig.-14c~~~~~~~~~~~~~~~~~\\
\includegraphics[scale=.5]{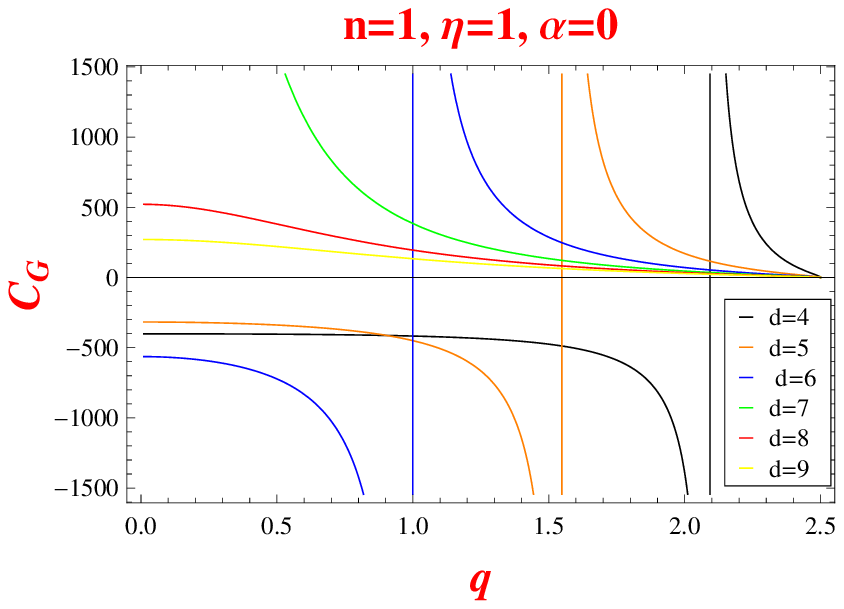}
\includegraphics[scale=.5]{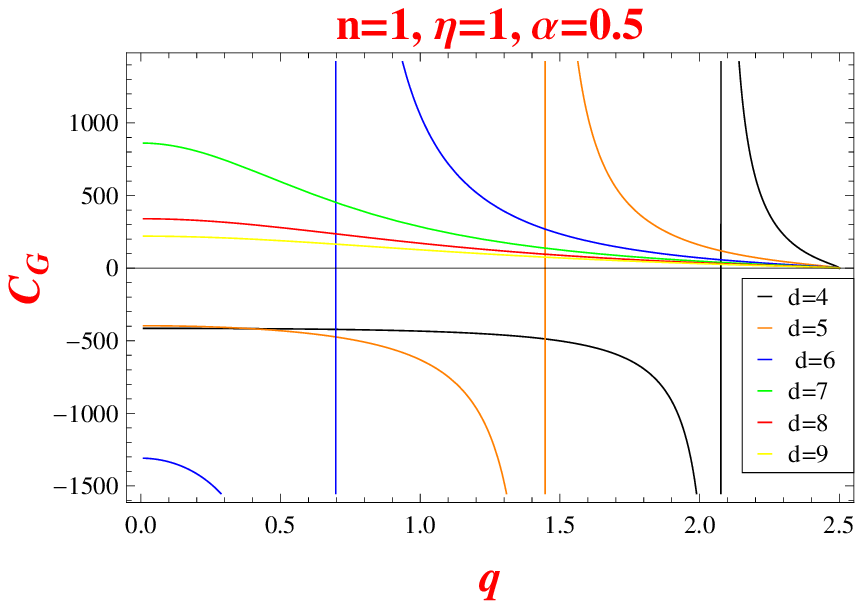}
\includegraphics[scale=.5]{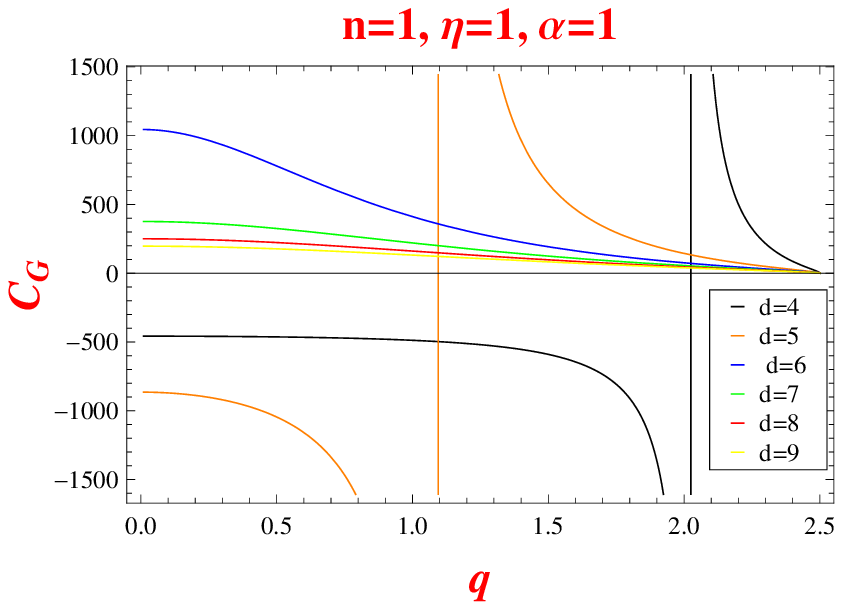}

Fig.-13a-13c represent the variations of $ C_G $ with respect to $ q $ in rainbow gravity with $\eta= 0.5$ for different dimensions $d$ and $\alpha=0, 0.5$ and $1$ and $m=2$.\\
Fig.-14a-14c represent the variations of $ C_G $ with respect to $ q $ in rainbow gravity with $\eta= 1$ for different dimensions $d$ and $\alpha=0, 0.5$ and $1$ and $m=2$.
\end{center} 
\end{figure}
When we plot the specific heat of the black hole incorporating the generalized uncertainty principle, we find for $\alpha=0$ the nature of the curves exactly coincides with that of the curves as depicted in Figures 3a to 4b. But when the curves are plotted for $\alpha=0.5$ and $1$, it is observed that the point of phase transitions shifts to the higher mass or higher charge region if $d$ is increased. Whereas $\alpha$ does not effect on the nature of the curves when we plot the $S_G$ vs $r_h$ or $S_G$ vs $q$ curves as shown in figures 11a to 14c.

From the similar way as above that taking $C_G=0$, the remnant mass of the black hole under GPU can be obtained as
\begin{equation}\label{ah9.equn23}
(m_{rem})_{G1}= \frac{\left(-l_p^2 \alpha ^2\right)^{\frac{1}{2}-\frac{d}{2}} \left(\left(-l_p^2 \alpha ^2\right)^d-l_p^6 q^2 \alpha ^6\right)}{l_p^4 \alpha ^4}
\end{equation}
and
\begin{equation}\label{ah9.equn24}
(m_{rem})_{G2}= \left(\frac{E_p^n}{\eta }\right)^{-\frac{d}{n}-\frac{3}{n}} \left(q^2 \left(\frac{E_p^n}{\eta }\right)^{\frac{2 d}{n}}+\left(\frac{E_p^n}{\eta}\right)^{6/n}\right).
\end{equation}
Thus we have found, two remnant masses under GPU and among them one is purely quantum corrected remnant mass expressed by equation (\ref{ah9.equn23}) and other is purely rainbow gravity inspired remnant mass given by equation (\ref{ah9.equn24}).

The entropy of rainbow gravity inspired  Reissner-Nordstr$\ddot{o}$m black hole in higher dimension under GUP is now computed keeping ${\cal O} (\eta^{3})$ and assuming $n\geq 3$ which reads as:
\begin{equation}\label{ah9.equn25}
S_G= \int \frac{dm}{\frac{d-3}{4\pi}\sqrt{1- \frac{\eta}{E_p^n} \frac{1}{r_h^n}} \left(\frac{m}{r_h^{d-2}} -\frac{2 q^2}{r_h^{2d-5}}\right)\left(1+\frac{\alpha^2 l_p^2}{2r_h^2}+... \right)^{-1}}
\end{equation}
which gives equation (\ref{ah9.equn35}) of Appendix for the quantum corrected entropy of the black hole.

An overall same trend is observed, if we incorporate generalized uncertainty principle (while we study the specific heat- figures 11a to 14c or entropy- figures 15a to 16c).  Existing's literature shows not only Einstein gravity but also Lovelock-Maxwell's gravity, etc. on their rainbow gravity counter part obeys first law of thermodynamics, i.e., $dM=Tds+ \Phi dQ$ \cite{Hendib-2017}. We will assume the charge to be constant through out and will find the entropy's expression.

\begin{figure}[h!]
\begin{center}
~~~~~~~~~Fig.-15a~~~~~~~~~~~~~~~~~~~~~~~~~Fig.-15b~~~~~~~~~~~~~~~~~~~~~~Fig.-15c~~~~~~~~~~~~~~~~~~\\
\includegraphics[scale=.5]{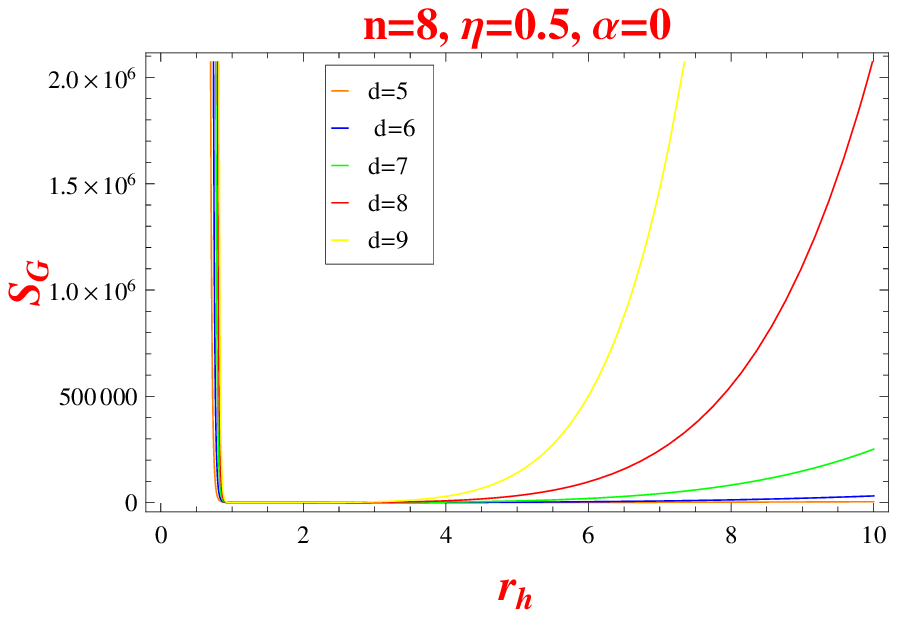}
\includegraphics[scale=.5]{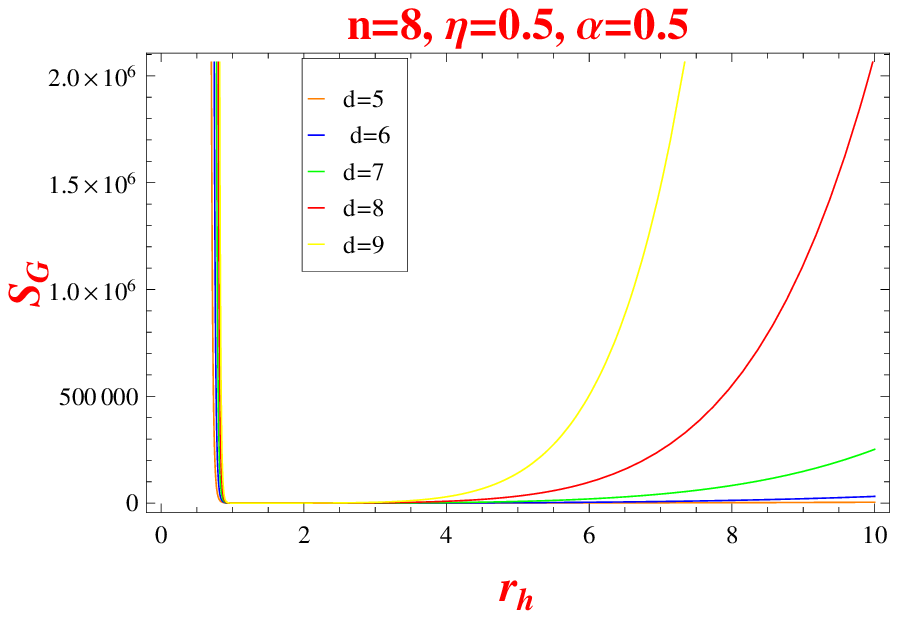}
\includegraphics[scale=.5]{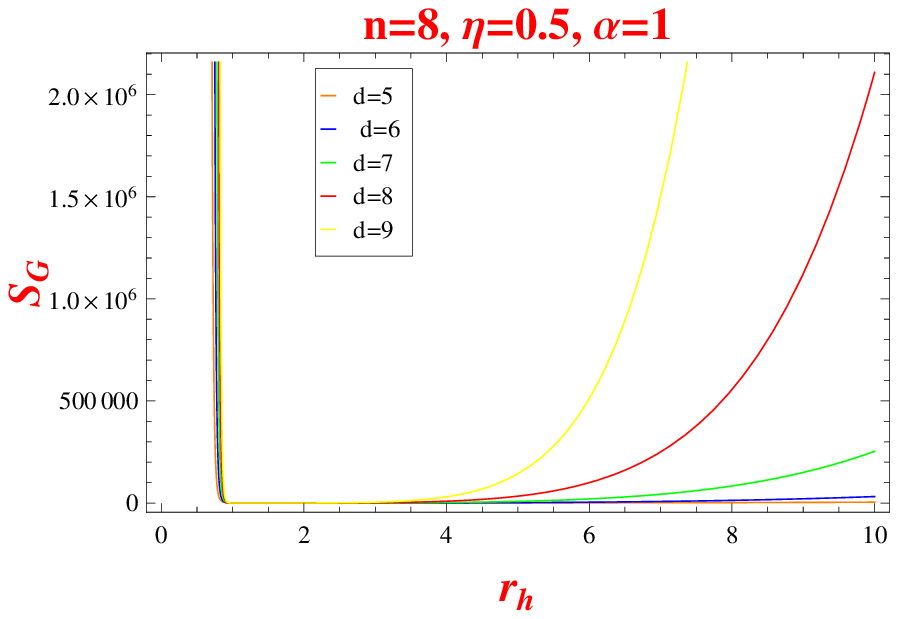}

~~~~~~~~Fig.-16a~~~~~~~~~~~~~~~~~~~~~~~~~Fig.-16b~~~~~~~~~~~~~~~~~~~~~~~Fig.-16c~~~~~~~~~~~~~~~~~\\
\includegraphics[scale=.5]{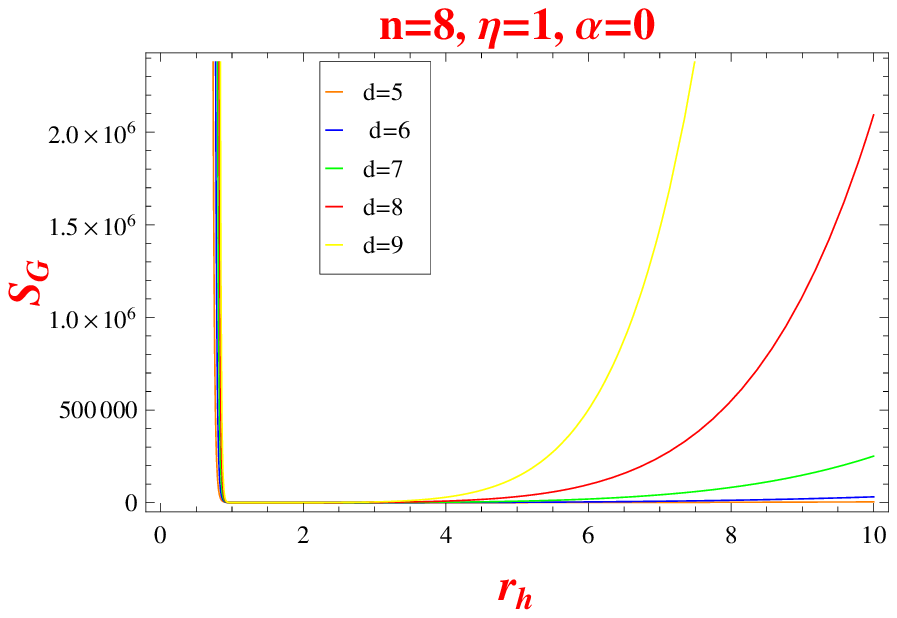}
\includegraphics[scale=.5]{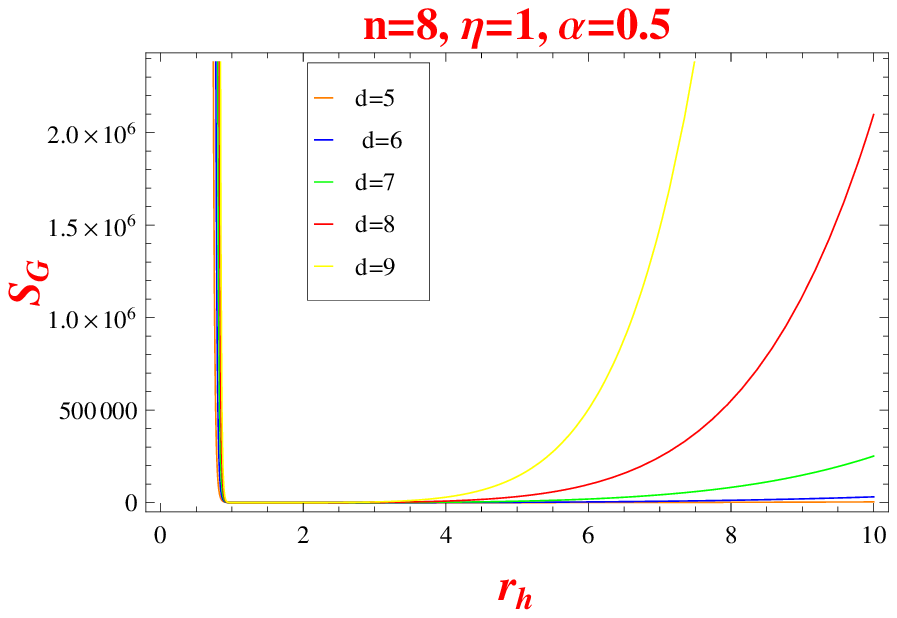}
\includegraphics[scale=.5]{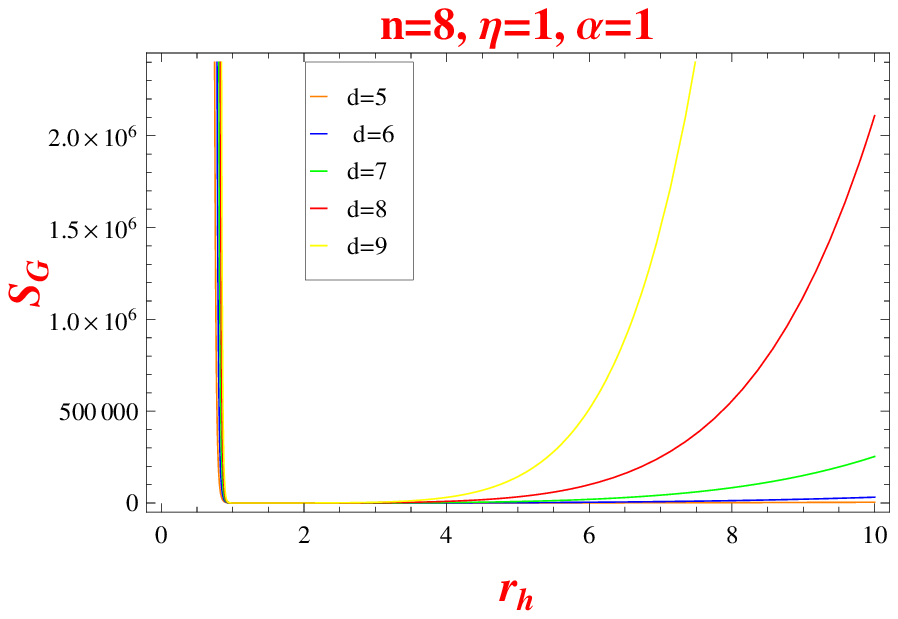}

Fig.-15a-15c represent the variations of $ S_G $ with respect to $ r_h $ in rainbow gravity with $\eta= 0.5$ for different dimensions $d$ and $\alpha=0, 0.5$ and $1$.\\
Fig.-16a-16c represent the variations of $ S_G $ with respect to $ r_h $ in rainbow gravity with $\eta= 1$ for different dimensions $d$ and $\alpha=0, 0.5$ and $1$.
\end{center} 
\end{figure}

\begin{figure}[h!]
\begin{center}
~~~~~~~~~Fig.-17a~~~~~~~~~~~~~~~~~~~~~~~~~Fig.-17b~~~~~~~~~~~~~~~~~~~~~~Fig.-17c~~~~~~~~~~~~~~~~~~\\
\includegraphics[scale=.5]{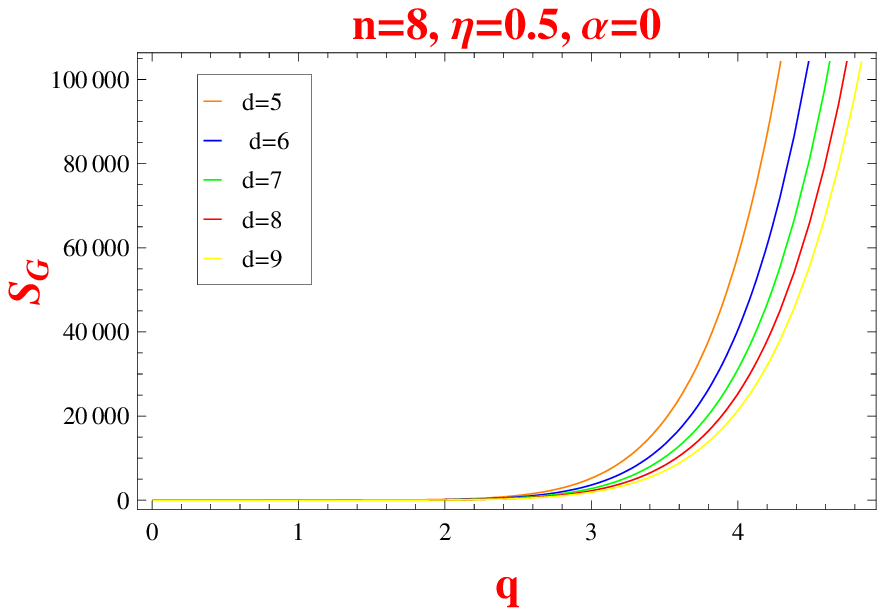}
\includegraphics[scale=.5]{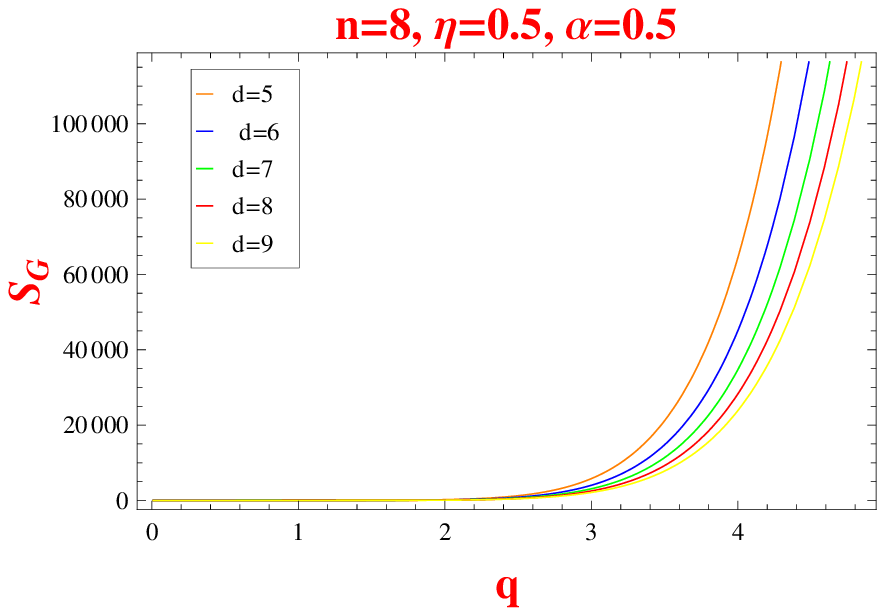}
\includegraphics[scale=.5]{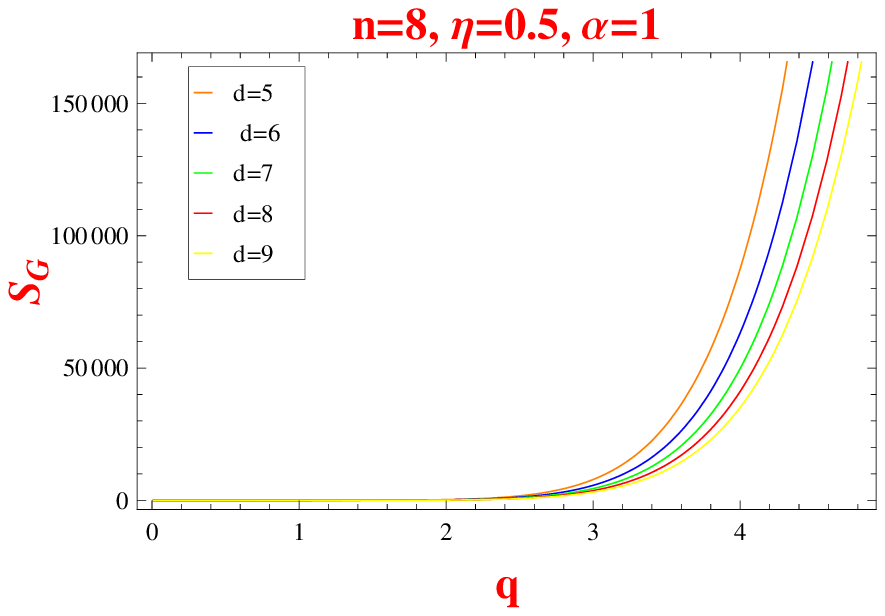}

~~~~~~~~Fig.-18a~~~~~~~~~~~~~~~~~~~~~~~~~Fig.-18b~~~~~~~~~~~~~~~~~~~~~~~Fig.-18c~~~~~~~~~~~~~~~~~\\
\includegraphics[scale=.5]{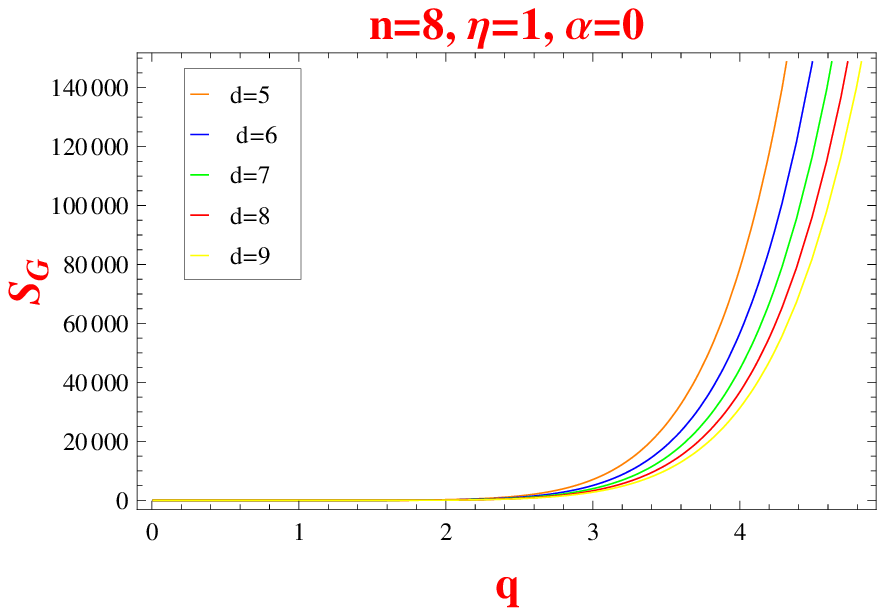}
\includegraphics[scale=.5]{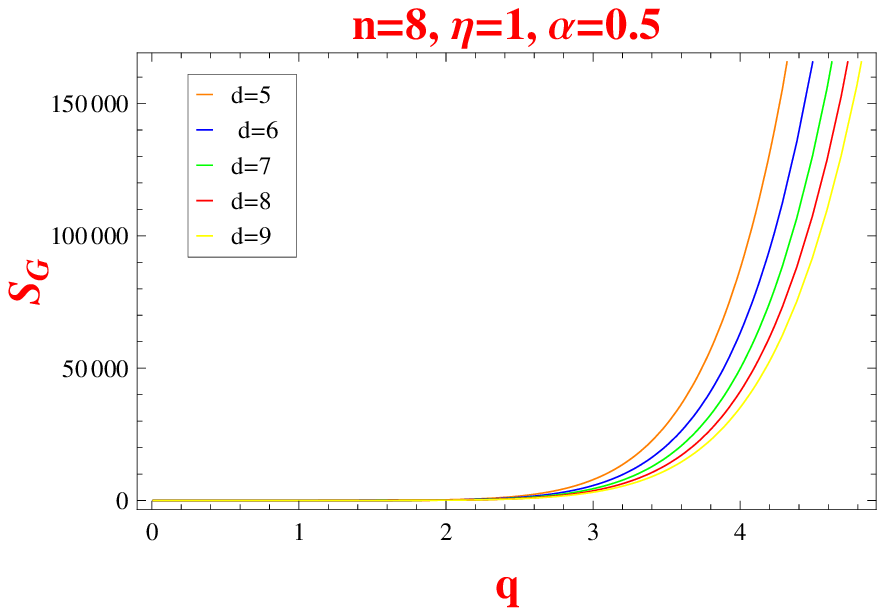}
\includegraphics[scale=.5]{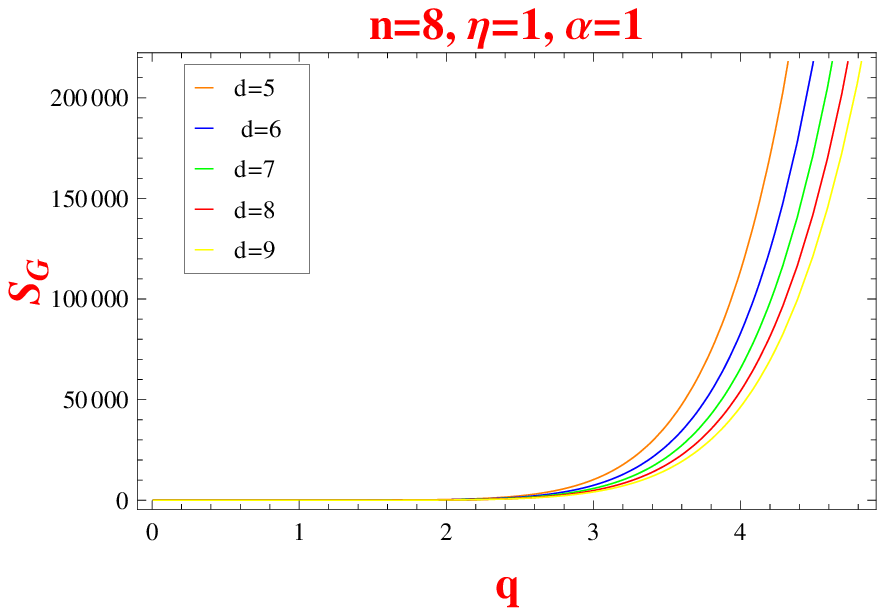}

Fig.-17a-17c represent the variations of $ S_G $ with respect to $ q $ in rainbow gravity with $\eta= 0.5$ for different dimensions $d$ and $\alpha=0, 0.5$ and $1$.\\
Fig.-18a-18c represent the variations of $ S_G $ with respect to $ q $ in rainbow gravity with $\eta= 1$ for different dimensions $d$ and $\alpha=0, 0.5$ and $1$.
\end{center} 
\end{figure}
\section{Phase Transition Under the Effect of Rainbow Gravity with General Uncertainty Principle}
To study the phase transitions in a higher dimensional Reissner-Nordstr$\ddot{o}$m black hole under the effect of rainbow gravity with General Uncertainty Principle, we assume a finite concentric spherical cavity in which the black hole is placed. The radius of the cavity is thought to be naturally larger than that of the black hole's horizon radius. The temperature which is fixed on the surface of the cavity, measured by the local observer, is called the local temperature and that is expressed as:
\begin{equation}\label{ah9.equn26}
T_{local}= \frac{T_G}{\sqrt{-g_{00}}}= \frac{\frac{d-3}{4\pi}\sqrt{1- \frac{\eta}{E_p^n} \frac{1}{r_h^n}} \left(\frac{m}{r_h^{d-2}} -\frac{2 q^2}{r_h^{2d-5}}\right)}{\left(1+\frac{\alpha^2 l_p^2}
{2 r_h^2}+... \right)\sqrt{1- \left(\frac{r_h}{r}\right)^{d-3}}}.
\end{equation}
In the limit $\alpha\rightarrow 0$ the equation (\ref{ah9.equn26}) becomes
\begin{equation}\label{ah9.equn27}
T_{local}= \frac{\frac{d-3}{4\pi}\sqrt{1- \frac{\eta}{E_p^n} \frac{1}{r_h^n}} \left(\frac{m}{r_h^{d-2}} -\frac{2 q^2}{r_h^{2d-5}}\right)}{\sqrt{1- \left(\frac{r_h}{r}\right)^{d-3}}},
\end{equation}
where $r_h$ is expressed in equation (\ref{ah9.equn4}).
From the definition of entropy and using the first law of entropy, one can compute the local energy as:

$$U_{local}= \int_{\left(r_h\right)_{rem}}^{r_{h}}T_{local} \times dS_G= \int_{\left(r_h\right)_{rem}}^{r_{h}}T_{local} \times \frac{dm}{T_G}$$
\begin{equation}\label{ah9.equn28}
= \left(r_h^{d-3}+ \frac{q^2}{r_h^{d-3}}\right) \left(\sqrt{1- \left(\frac{\left(r_h\right)_{rem}}{r}\right)^{d-3}}- \sqrt{1- \left(\frac{r_h}{r}\right)^{d-3}}\right),
\end{equation} 
where $\left(r_h\right)_{rem}$ is the remnant temperature of the black hole and is expressed as:
\begin{equation}\label{ah9.equn29}
\left(r_h\right)_{rem}=\left( \frac{(m_{rem})_{G1}}{2}+ \frac{\sqrt{(m_{rem})_{G1}^2-4 q^2} }{2}\right)^{\frac{1}{d-3}}
\end{equation}

\begin{figure}[h!]
\begin{center}
~~~~~~~~~Fig.-19a~~~~~~~~~~~~~~~~~~~~~~~~~Fig.-19b~~~~~~~~~~~~~~~~~~~~~~Fig.-19c~~~~~~~~~~~~~~~~~~\\
\includegraphics[scale=.5]{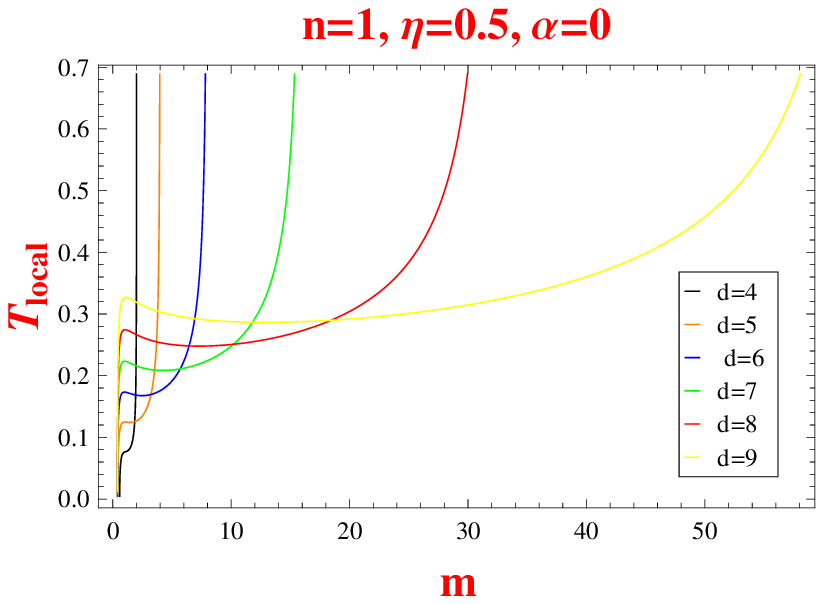}
\includegraphics[scale=.5]{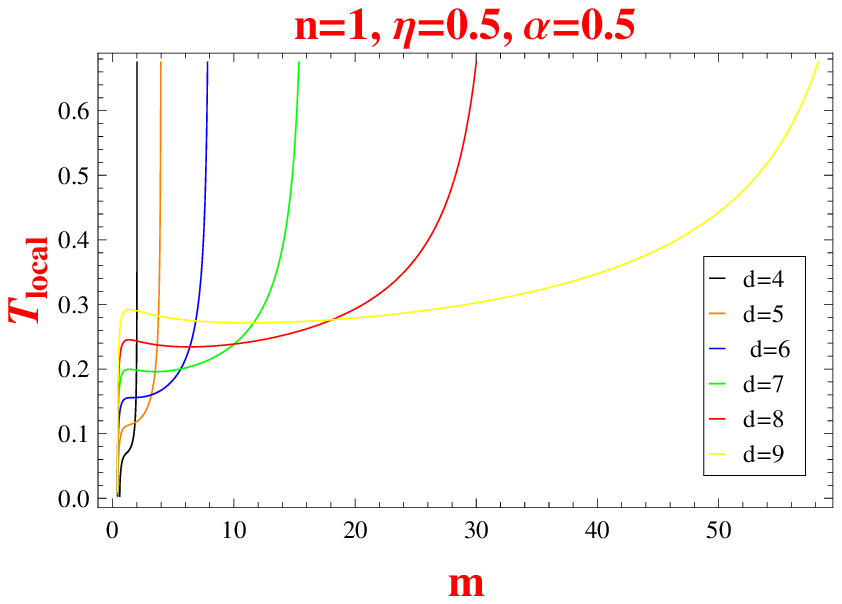}
\includegraphics[scale=.5]{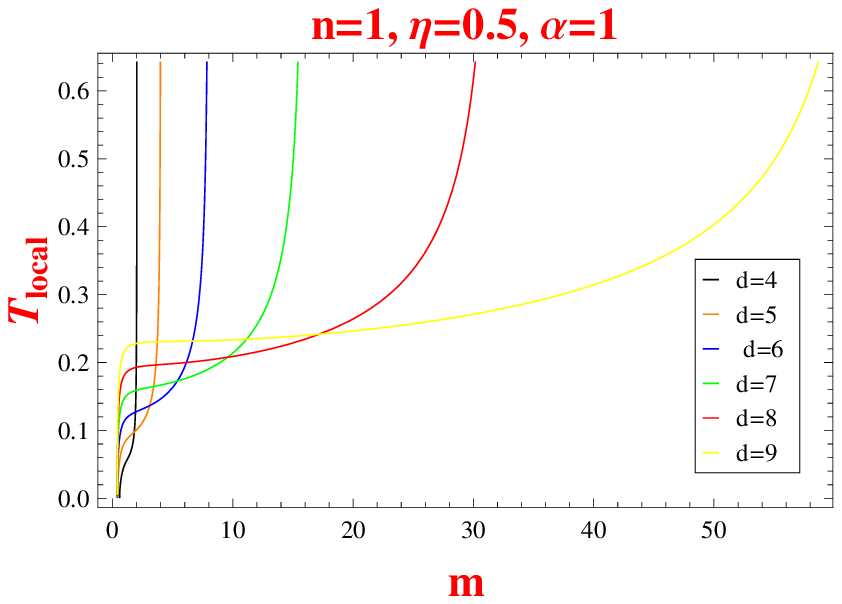}

~~~~~~~~Fig.-20a~~~~~~~~~~~~~~~~~~~~~~~~~Fig.-20b~~~~~~~~~~~~~~~~~~~~~~~Fig.-20c~~~~~~~~~~~~~~~~~\\
\includegraphics[scale=.5]{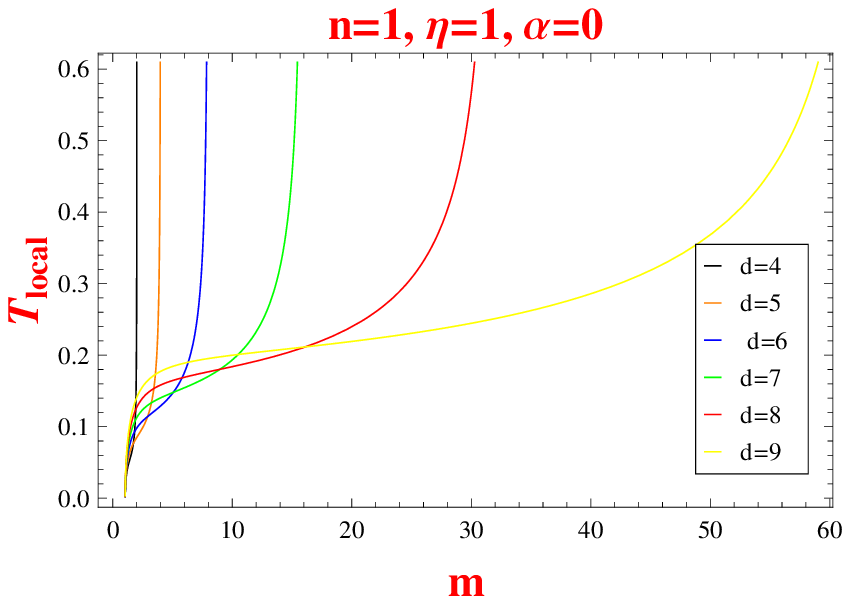}
\includegraphics[scale=.5]{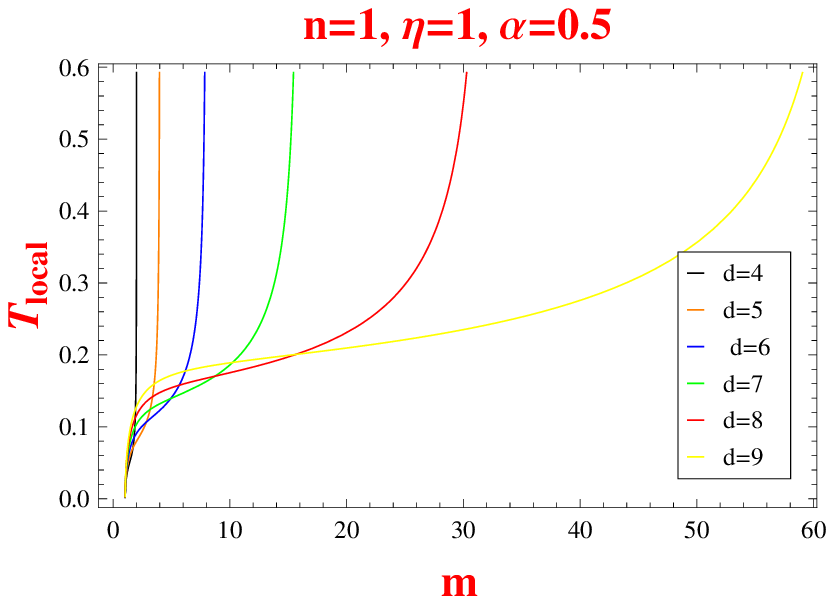}
\includegraphics[scale=.5]{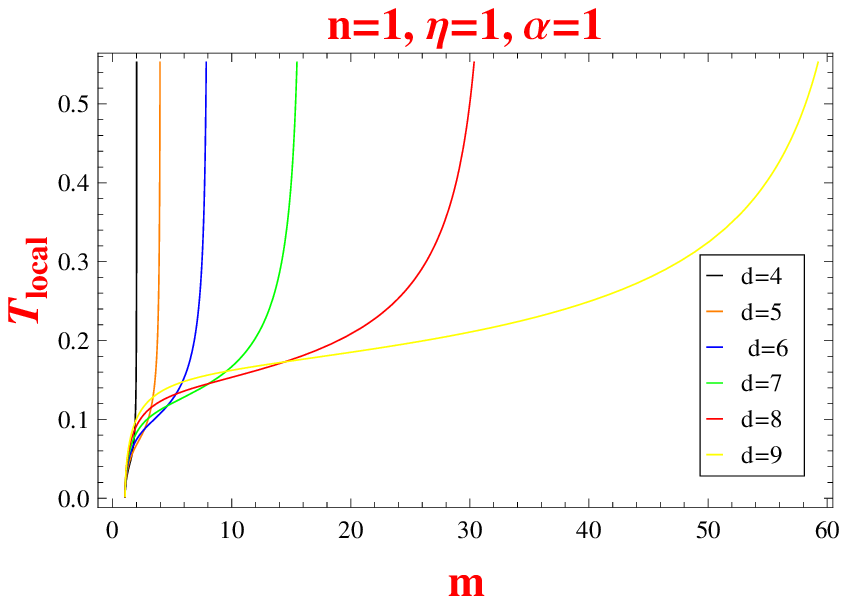}

Fig.-19a-19c represent the variations of $ T_{local} $ with respect to $ m $ in rainbow gravity with $\eta= 0.5$ for different dimensions $d$ and $\alpha=0, 0.5$ and $1$.\\
Fig.-20a-20c represent the variations of $ T_{local} $ with respect to $ m $ in rainbow gravity with $\eta= 1$ for different dimensions $d$ and $\alpha=0, 0.5$ and $1$.
\end{center} 
\end{figure}

\begin{figure}[h!]
\begin{center}
~~~~~~~~~Fig.-21a~~~~~~~~~~~~~~~~~~~~~~~~~Fig.-21b~~~~~~~~~~~~~~~~~~~~~~Fig.-21c~~~~~~~~~~~~~~~~~~\\
\includegraphics[scale=.5]{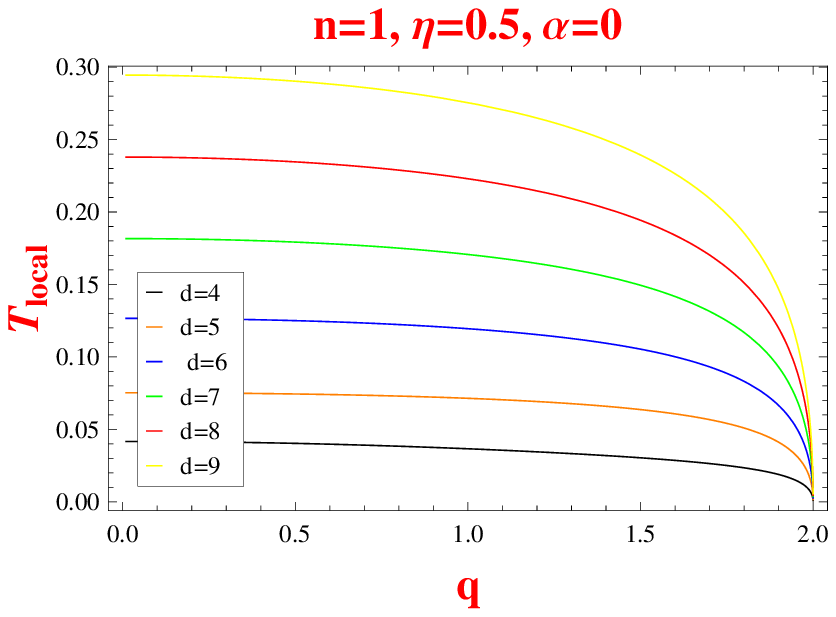}
\includegraphics[scale=.5]{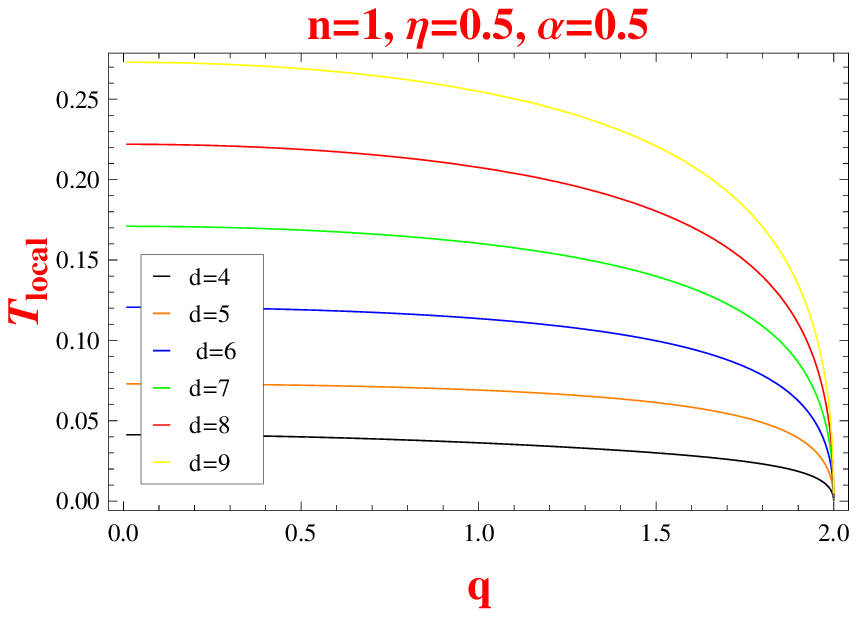}
\includegraphics[scale=.5]{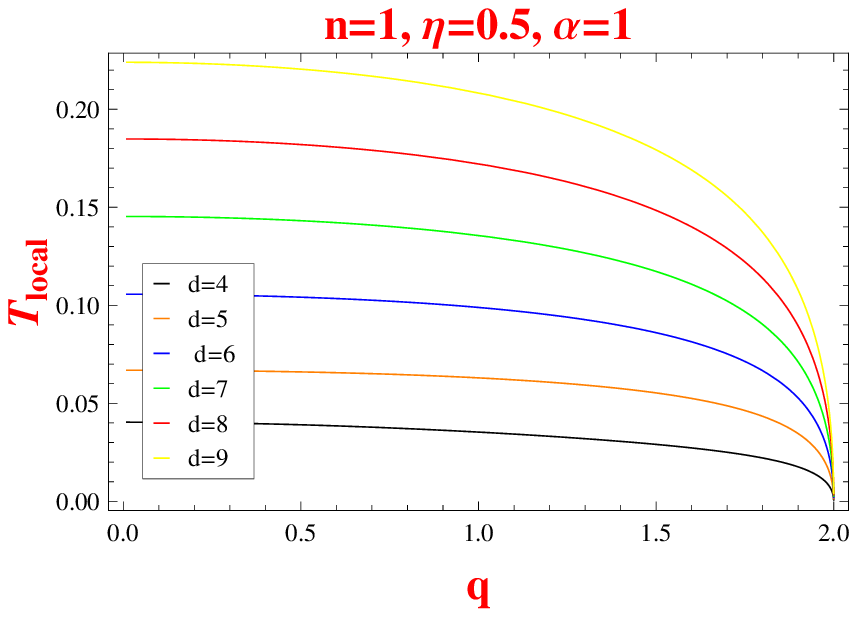}

~~~~~~~~Fig.-22a~~~~~~~~~~~~~~~~~~~~~~~~~Fig.-22b~~~~~~~~~~~~~~~~~~~~~~~Fig.-22c~~~~~~~~~~~~~~~~~\\
\includegraphics[scale=.5]{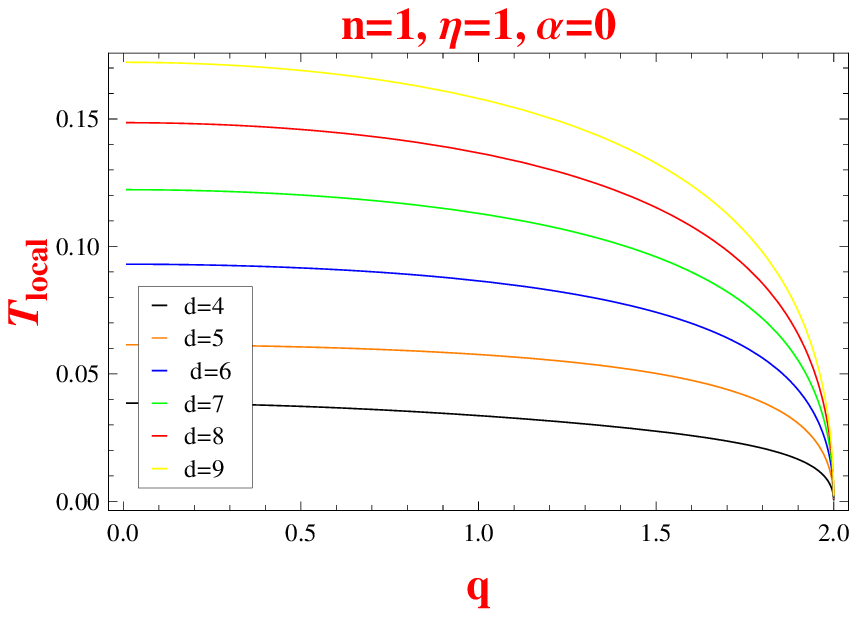}
\includegraphics[scale=.5]{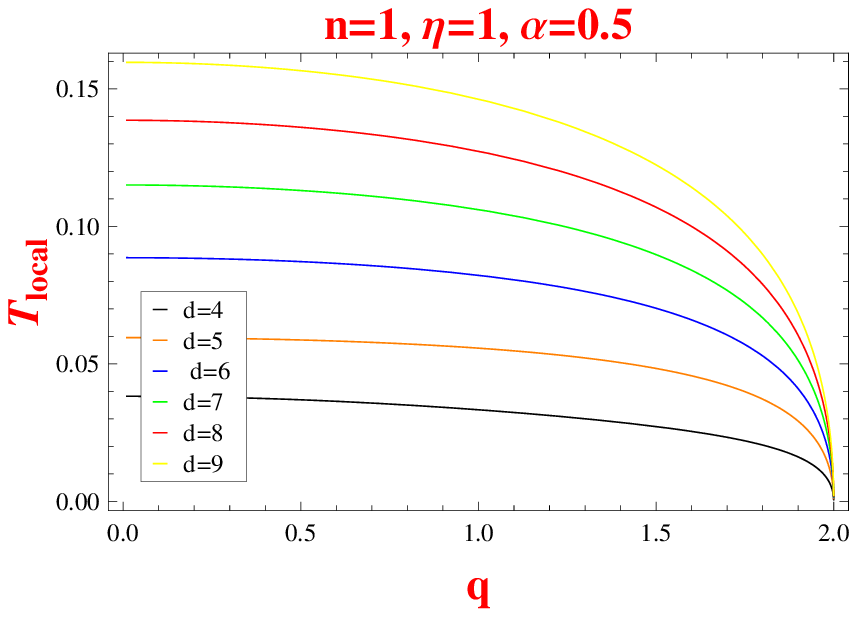}
\includegraphics[scale=.5]{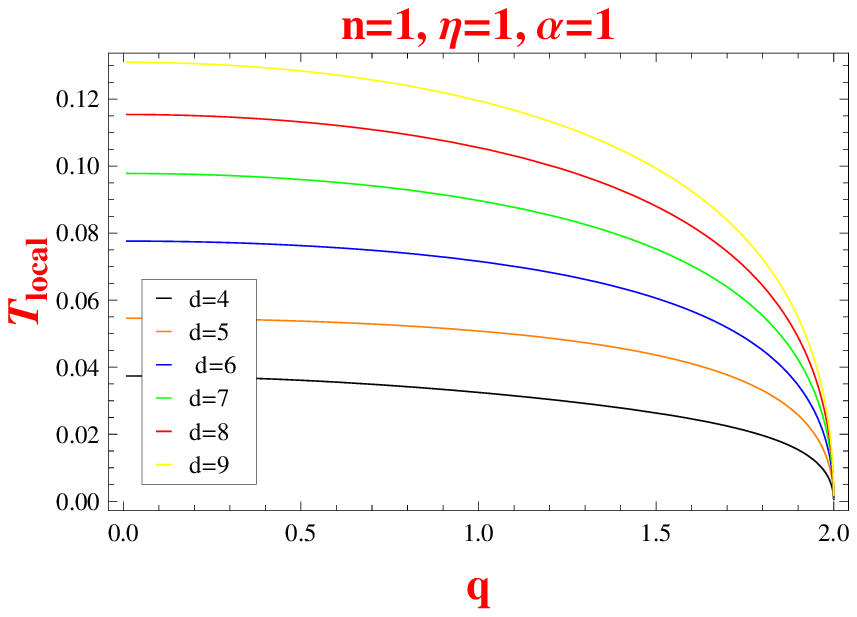}

Fig.-21a-21c represent the variations of $ T_{local} $ with respect to $ q $ in rainbow gravity with $\eta= 0.5$ for different dimensions $d$ and $\alpha=0, 0.5$ and $1$.\\
Fig.-22a-22c represent the variations of $ T_{local} $ with respect to $ q $ in rainbow gravity with $\eta= 1$ for different dimensions $d$ and $\alpha=0, 0.5$ and $1$.
\end{center} 
\end{figure}
We plot the variations of $T_{local}$ with respect to $q$ for $\eta=0.5, 1$ and $\alpha=0, 0.5, 1$ and for different dimensions $d$ in figures 19a to 20c. For $\eta=0.5$, we observe for low $m$ there is a local maxima and for slight increment of $m$ we also find a local minima that means there are two phase transitions. For further increment of $m$, the value of $T_{local}$ increases gradually. We also find from the curves that for higher values of $d$ the position of the local maxima is unaltered but the local minima shifts to higher mass region. The nature of the curves does not depend on the GUP parameter $\alpha$. For $\eta=1$, the sharpness of the peak of the local maxima reduces.

When we plot the same, taking $m$ constant with varying $q$ (Figures 21a to 22c), we notice for low $q$, all the curves are parallel. When we increase $q$, $T_{local}$ reduces and after a while different $T_{local}$ graphs converge to a point for all $d$ and whatever be the values of $\eta$ and $\alpha$.

The thermal heat capacity specifies the thermal stability of the black holes. Therefore to examine the thermal stability for this black hole under rainbow gravity back ground, one can compute the local heat capacity as:
\begin{equation}\label{ah9.equn30}
C_{local}= \frac{\partial U_{local}}{\partial T_{local}}= \frac{X(r_h, r, \eta, E_p)}{Y(r_h, r, \eta, E_p)},
\end{equation}
where $$X(r_h, r, \eta, E_p)= 8 \pi  E_p^n r_h^{d+n-3} \left(1-\left(\frac{r_h}{r}\right)^{d-3}\right)^{3/2} \left(\frac{l_p^2 \alpha ^2}{2 r_h^2}+1\right)^2 \sqrt{1-\eta  r_h^{-n} E_p^{-n}} \left(2 r_h^3 \left(r_h^{2 d}-q^2 r_h^6\right)\right.$$

$$ \left(\sqrt{1-\left(\frac{\left(r_h\right)_{rem}}{r_h}\right)^{d-3}}-\sqrt{1-\left(\frac{r_h}{r}\right)^{d-3}}\right)+\frac{x^3 \left(r_h^{2 d}+q^2 r_h^6\right) \left(\frac{r_h}{r}\right)^d}{\sqrt{1-\left(\frac{r_h}{r}\right)^{d-3}}}\bigg)$$ and
   
$$Y(r_h, r, \eta, E_p)= 4 l_p^2 \alpha ^2 \left(r_h^3-r^3 \left(\frac{r_h}{r}\right)^d\right) \left(m r_h^d-2 q^2 r_h^3\right) \left(r_h^n E_p^n-\eta \right)+n \eta  \left(r_h^3-r^3 \left(\frac{r_h}{r}\right)^d\right) \left(l_p^2 \alpha ^2+2 r_h^2\right) \left(m r_h^d-2 q^2 r_h^3\right)$$
   
$$+(d-3) r^3 \left(\frac{r_h}{r}\right)^d \left(l_p^2 \alpha ^2+2 r_h^2\right) \left(m r_h^d-2 q^2 r_h^3\right) \left(r_h^n E_p^n-\eta \right)-2 \left(r_h^3-r^3 \left(\frac{r_h}{r}\right)^d\right) \left(l_p^2 \alpha ^2+2 r_h^2\right) \left((d-2) m r_h^d\right.$$

$$+2 (5-2 d) q^2 r_h^3 \bigg) \left(r_h^n E_p^n-\eta \right)$$

\begin{figure}[h!]
\begin{center}
~~~~~~~~~Fig.-23a~~~~~~~~~~~~~~~~~~~~~~~~~Fig.-23b~~~~~~~~~~~~~~~~~~~~~~Fig.-23c~~~~~~~~~~~~~~~~~~\\
\includegraphics[scale=.5]{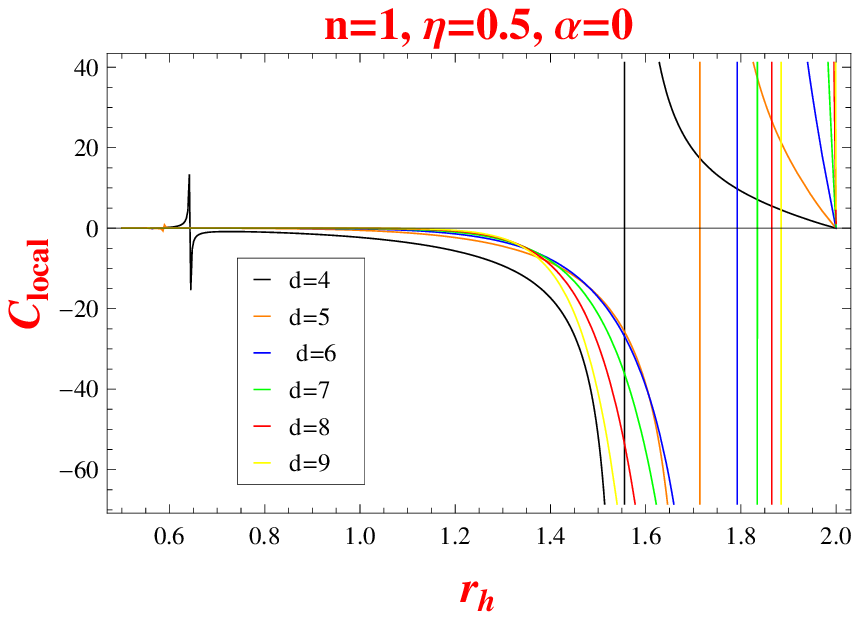}
\includegraphics[scale=.5]{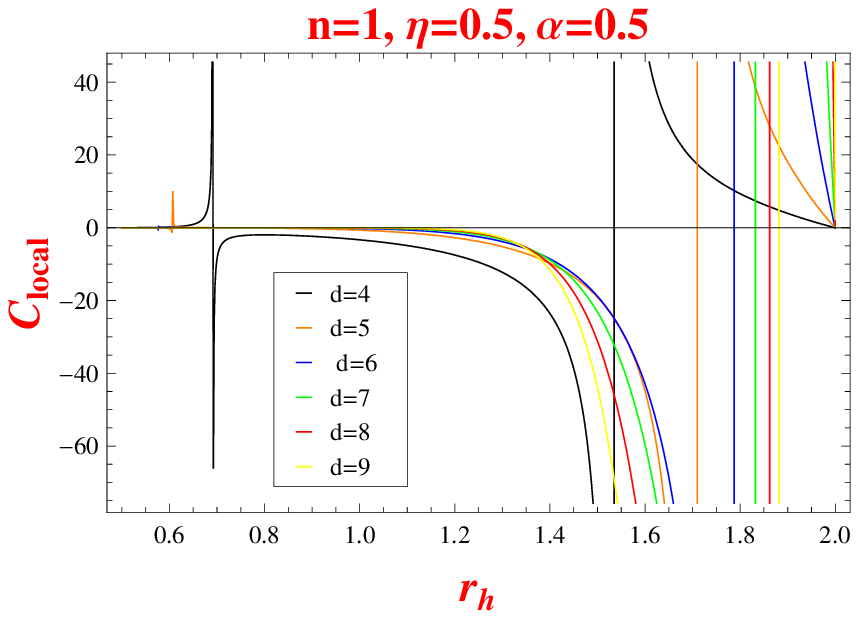}
\includegraphics[scale=.5]{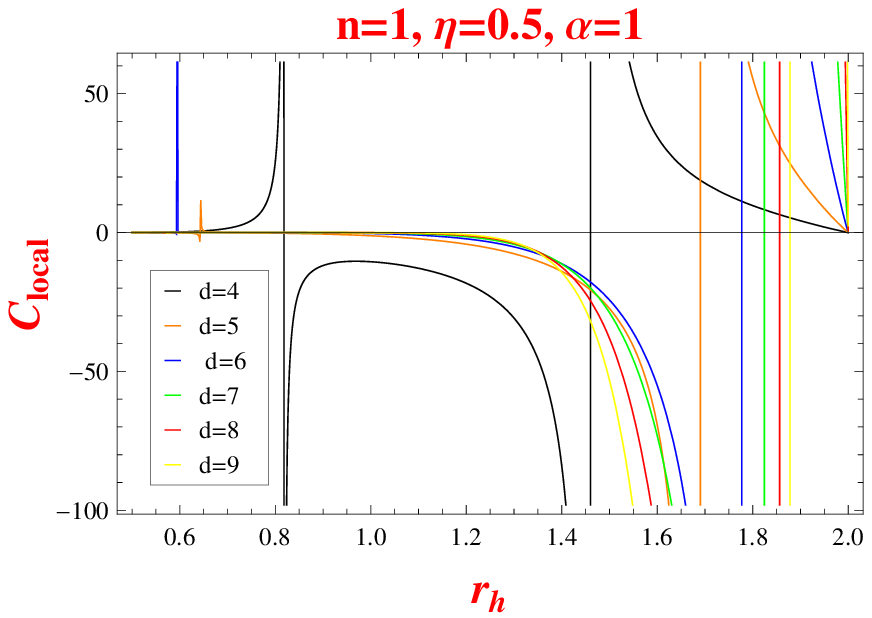}

~~~~~~~~Fig.-24a~~~~~~~~~~~~~~~~~~~~~~~~~Fig.-24b~~~~~~~~~~~~~~~~~~~~~~~Fig.-24c~~~~~~~~~~~~~~~~~\\
\includegraphics[scale=.5]{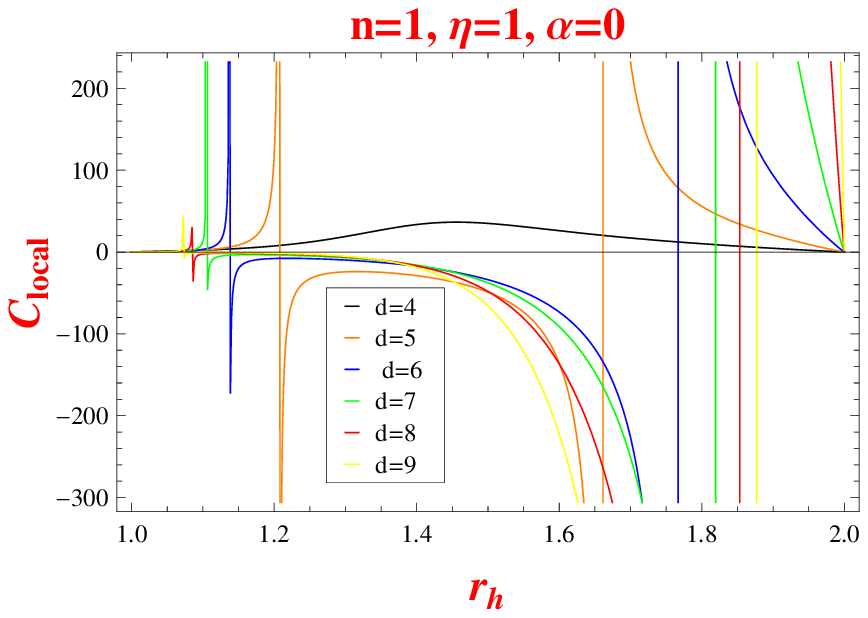}
\includegraphics[scale=.5]{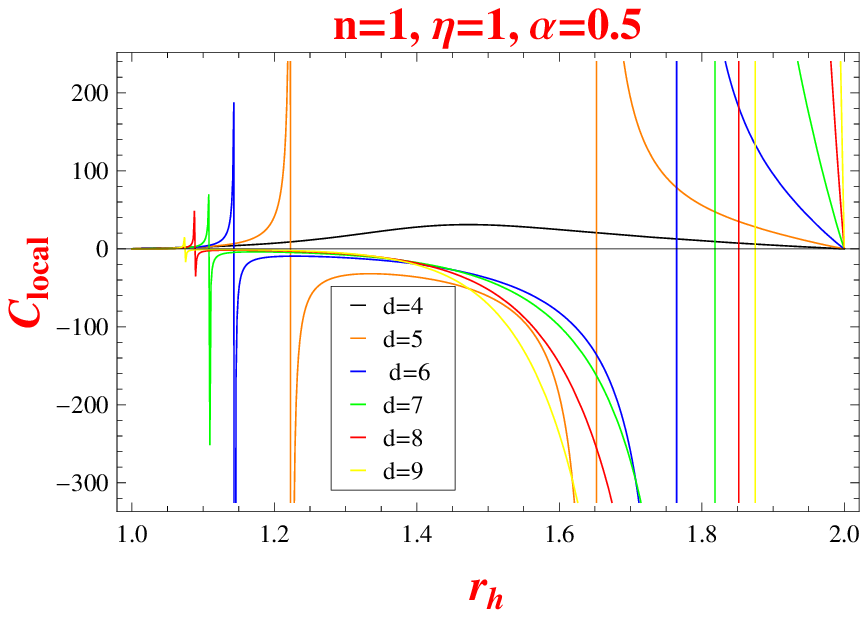}
\includegraphics[scale=.5]{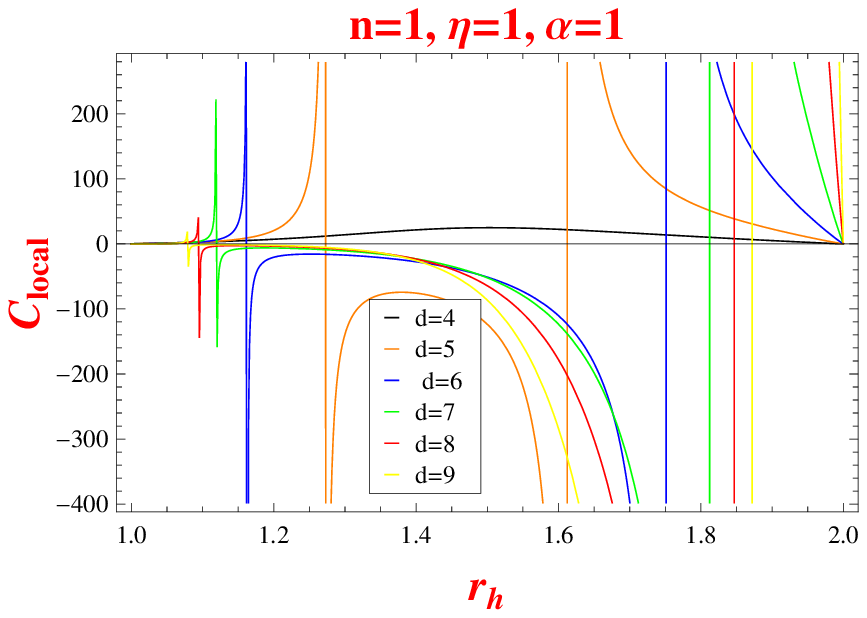}

Fig.-23a-23c represent the variations of $ C_{local} $ with respect to $ r_h $ in rainbow gravity with $\eta= 0.5$ for different dimensions $d$ and $\alpha=0, 0.5$ and $1$.\\
Fig.-24a-24c represent the variations of $ C_{local} $ with respect to $ r_h $ in rainbow gravity with $\eta= 1$ for different dimensions $d$ and $\alpha=0, 0.5$ and $1$.
\end{center} 
\end{figure}
In figures 23a to 24c, we depict $C_{local}$ vs $r_h$ with the variations of $\eta$ and $\alpha$ and for different $d$. When we take $\eta=0.5$ and vary the values of $\alpha$, we get the curves having almost equal nature where at low horizon region we notice a sharp maxima and a sharp minima for all $d$. The physical meaning of this behavior of the curves is that there is a phase transition at low horizon region. Here the phase transition transits a stable black hole to its unstable counter part. Again there is another phase transition which occurs at high horizon region but here the phase transition transits a unstable black hole to its stable counter part. For the increment of $\alpha$, the peak of the maxima or minima increases and for the increment of $d$ the transition points shift towards the high horizon region.

When we do the same for $\eta=1$, we observe that the overall natures of the curves are same and the only difference is that the peaks, i.e., the phase transition points shift towards the high horizon region.  
\section{Brief Discussions and Conclusions}
In this paper, we have investigated the thermodynamic properties of rainbow inspired Reissner-Nordstr$\ddot{o}$m black hole in higher dimensions $d$ incorporating the effect of the generalized uncertainty principle. Firstly we have established the relation between the black hole temperature $T_h$ at the event horizon, mass $m$ and charge $q$ parameters. Here we have observed that the rainbow gravity plays crucial role to modify this mass-temperature relation. We have also found that the insertion of rainbow gravity to black hole solutions leads to the existence of a black hole remnant. For rainbow inspired Reissner-Nordstr$\ddot{o}$m black hole, we have noticed that the remnant mass and the critical mass are equal and possess only one value, whereas under general uncertainty principle effects two values of remnant mass are found : one of them resembled with the previous value which does not depend on generalized uncertainty principle parameter $\alpha$ and other involved with $\alpha$ and Planck length $l_p$. This may be the new result which has been computed here. For rainbow inspired Schwarzchild black hole in higher dimensions the remnant mass does not depend on $\eta$. It depends only on $\alpha$ and has only one value under the effects of generalized uncertainty principle \cite{Mandal-2018} unlike our work. Here we depict the thermodynamic properties like temperature and entropy for this black hole in the rainbow gravity framework separately as well as under the effects of generalized uncertainty principle simultaneously for various values of $\eta$, $\alpha$ and for different $d$. But we did not find remarkable differences between the curves, which means the generalized uncertainty principle does not effect significantly on those properties. We have also calculated temperature, internal energy and heat capacity for the black hole introducing the concept of local observer and have examined the thermal stability. Again by plotting the $C_{local}$ vs $r_h$ curves, we have analyzed the phase transitions. Here we have found that there are two local maxima and two local minima at the low horizon region and at the large horizon region which signifies that there are two phase transitions at the horizons. We have also observed here that for increment of $d$, the transition point shifts towards the high horizon regions.

\vspace{.1 in}
{\bf Acknowledgment:}
This research is supported by the project grant of Goverment of West Bengal, Department of Higher Education, Science and Technology and Biotechnology (File no:- $ST/P/S\&T/16G-19/2017$). AH wishes to thank the Department of Mathematics, the University of Burdwan for the research facilities provided during the work. RB thanks IUCAA, PUNE for Visiting Associateship.

\section{Appendix}
The heat capacity of the black hole is given as:
$$C_h=\left[\pi  \sqrt{m^2-4 q^2} \sqrt{1- \frac{\eta}{E_p^n} \left(\frac{1}{2} \sqrt{m^2-4 q^2}+\frac{m}{2}\right)^{-\frac{n}{d-3}}}\right] \left[2^{-\frac{d-4}{d-3}} \left(\left(2^{\frac{d}{d-3}+1} (2 d-5) q^2 \left(\frac{1}{2} \sqrt{m^2-4 q^2}+\frac{m}{2}\bigg)^{\frac{3}{d-3}}\right. \right. \right. \right.$$

$$+\left(\sqrt{m^2-4 q^2}+m\bigg)^{\frac{d}{d-3}} \left((d-3) \sqrt{m^2-4 q^2}+(2-d) m\right)\right) \left(\sqrt{m^2-4 q^2}+m\right)^{-\frac{2(d-1)}{d-3}} $$

$$\left(1- \frac{\eta}{E_p^n} \left(\frac{1}{2} \sqrt{m^2-4 q^2}+\frac{m}{2}\right)^{-\frac{n}{d-3}}\right)+2^{\frac{n}{d-3}-1} \frac{1}{E_p^n} \left(m n \eta  \left(\sqrt{m^2-4 q^2}+m\right)^{-\frac{d+n-2}{d-3}}\right.$$
     
\begin{equation}\label{ah9.equn31}
-4 n q^2 \eta \left(\sqrt{m^2-4 q^2}+m\right)^{-\frac{2 d+n-5}{d-3}}\bigg)\bigg)\bigg]^{-1}.
\end{equation}

The entropy of the black hole is expressed as:
$$S_h=\frac{1}{4} \pi  E_p^{-3 n} \left(r_h^{-7 d-n} \left(-\frac{8 q^8 r_h^{22} \eta  E_p^{2 n}}{-7 d-n+22}+\frac{8 q^6 \eta  r_h^{2 d+16} E_p^{2 n}}{-5 d-n+16}-\frac{8 q^4 \eta  r_h^{4 d+10} E_p^{2 n}}{-3
   d-n+10}+r_h^{2 (d+5)} \left(\frac{8 q^6 r_h^6 \eta  E_p^{2 n}}{-5 d-n+16}\right. \right. \right.$$
   
$$+\frac{8 q^4 \eta  r_h^{2 d} E_p^{2 n}}{3 d+n-10}\bigg)+\frac{8 \eta  r_h^{8 d-2} E_p^{2 n}}{d-n-2}\bigg)+r_h^{-7 d-2 n} \left(-\frac{6 q^8 r_h^{22} \eta ^2 E_p^n}{-7 d-2 n+22}+\frac{6 q^6 \eta ^2 r_h^{2 d+16} E_p^n}{-5 d-2 n+16}-\frac{6 q^4 \eta ^2 r_h^{4 d+10} E_p^n}{-3 d-2 n+10}+r_h^{2 (d+5)}\right.$$

$$\left(\frac{6 q^6 r_h^6 \eta ^2 E_p^n}{-5 d-2 n+16}+\frac{6 q^4 \eta ^2 r_h^{2 d} E_p^n}{3 d+2 n-10}\right)+\frac{6 \eta ^2 r_h^{8 d-2} E_p^n}{d-2 n-2}\bigg)+r_h^{-7 d-3 n} \left(-\frac{5 q^8 r_h^{22} \eta ^3}{-7 d-3 n+22}+\frac{5 q^6 \eta ^3 r_h^{2 d+16}}{-5 d-3 n+16}\right.$$

$$-\frac{5 q^4 \eta ^3 r_h^{4 d+10}}{10-3 (d+n)}+r_h^{2 (d+5)} \left(\frac{5 q^6 r_h^6 \eta ^3}{-5 d-3 n+16}+\frac{5 q^4 \eta ^3 r_h^{2 d}}{3 d+3 n-10}\right)+\frac{5 \eta ^3 r_h^{8 d-2}}{d-3 n-2}\bigg)+r_h^{-7 d} \left(-\frac{16 q^8 r_h^{20} E_p^{3 n}}{20-7 d}\right.$$

$$ +\frac{16 q^6 r_h^{2 d+14} E_p^{3 n}}{14-5 d}-\frac{16 q^4 r_h^{4 d+8} E_p^{3
   n}}{8-3 d}+r_h^{6 d} \left(\frac{16 q^2 r_h^2 E_p^{3 n}}{2-d}-\frac{16 q^2 r_h^4 E_p^{3 n}}{4-d}\right) $$

\begin{equation}\label{ah9.equn32}
+r_h^{2 (d+5)} \left(\frac{16 q^6 r_h^4 E_p^{3 n}}{4-5 (d-2)}+\frac{16 q^4 r_h^{2 d-2} E_p^{3 n}}{3 d-8}\right)+\frac{16 r_h^{8 d-2} E_p^{3 n}}{d-2}\bigg)\bigg).
\end{equation}

The quantum corrected temperature of the black hole is given by
\begin{equation}\label{ah9.equn33}
T_G=\frac{d-3}{4\pi}\sqrt{1- \frac{\eta}{E_p^n} \frac{1}{r_h^n}} \times \left\{\frac{m}{\left(\frac{m}{2} + \frac{\sqrt{m^2-4q^2}}{2} \right)^{\frac{d-2}{d-3}}} -\frac{2 q^2}{\left(\frac{m}{2} + \frac{\sqrt{m^2-4q^2}}{2} \right)^{\frac{{2d-5}}{d-3}}}\right\} \left\{1+\frac{\alpha^2 l_p^2}
{\left(\frac{m}{2} + \frac{\sqrt{m^2-4q^2}}{2} \right)^{\frac{2}{d-3}}}+... \right\}^{-1}
\end{equation}

The quantum corrected heat capacity of the black hole is given by
$$C_G =\left( \pi  \sqrt{m^2-4 q^2} \left(\left(\frac{1}{2} \sqrt{m^2-4 q^2}+\frac{m}{2}\right)^{\frac{2}{d-3}}+l_p^2 \alpha ^2\right)^2 \sqrt{1- \frac{\eta}{E_p^n} \left(\frac{1}{2} \sqrt{m^2-4 q^2}+\frac{m}{2}\right)^{-\frac{n}{d-3}}}\right)$$
   
$$\left[\left(2^{-\frac{4d+1}{d-3}}\right) \left(\sqrt{m^2-4 q^2}+m\right)^{-\frac{2(d-2)}{d-3}} \left(n q^2 \frac{\eta}{E_p^n} \left(-2^{\frac{4 d+n}{d-3}}\right) \left(\sqrt{m^2-4 q^2}+m\right)^{\frac{3-n}{d-3}}\right. \right.$$
   
$$\left(\left(\frac{1}{2} \sqrt{m^2-4 q^2}+\frac{m}{2}\right)^{\frac{2}{d-3}}+l_p^2 \alpha ^2\right)+m n \frac{\eta}{E_p^n} \left( 2^{\frac{2 d+n+6}{d-3}} \right) \left(\sqrt{m^2-4 q^2}+m\right)^{\frac{d-n}{d-3}} \left(\left(\frac{1}{2} \sqrt{m^2-4 q^2}+\frac{m}{2}\right)^{\frac{2}{d-3}}+l_p^2 \alpha ^2\right)$$
   
$$+\left(32^{\frac{d}{d-3}}\right) l_p^2  \alpha ^2 \left(m \left(\frac{1}{2} \sqrt{m^2-4 q^2}+\frac{m}{2}\right)^{\frac{d}{d-3}}-2 q^2 \left(\frac{1}{2} \sqrt{m^2-4 q^2}+\frac{m}{2}\right)^{\frac{3}{d-3}}\right)
 \left(1-\frac{\eta}{E_p^n} \left(\frac{1}{2} \sqrt{m^2-4 q^2}+\frac{m}{2}\right)^{-\frac{n}{d-3}}\right) $$

$$+8^{\frac{d+1}{d-3}} \left(\left(2^{\frac{2d-3}{d-3}}\right) (2 d-5) q^2 \left(\frac{1}{2}
   \sqrt{m^2-4 q^2}+\frac{m}{2}\right)^{\frac{3}{d-3}}+\left(\sqrt{m^2-4 q^2}+m\right)^{\frac{d}{d-3}} \left((d-3) \sqrt{m^2-4 q^2}+(2-d) m\right)\right) $$   

\begin{equation}\label{ah9.equn34}
\left(\left(\frac{1}{2} \sqrt{m^2-4 q^2}+\frac{m}{2}\right)^{\frac{2}{d-3}}+l_p^2 \alpha ^2\right) \left(1-\frac{\eta}{E_p^n} \left(\frac{1}{2} \sqrt{m^2-4 q^2}+\frac{m}{2}\right)^{-\frac{n}{d-3}}\right)\bigg)\bigg]^{-1}.
\end{equation}

The quantum corrected entropy of the black hole is obtained as:
$$S_G= \frac{1}{8} E_p^{-3 n} \left(\left(\left(\frac{32 q^6 r_h^{14} E_p^{3 n}}{14-5 d}+\frac{16 l_p^2 q^6 r_h^{12} \alpha ^2 E_p^{3 n}}{12-5 d}\right) r_h^{2 d}+\left(-\frac{32 q^4 r_h^8 E_p^{3 n}}{8-3 d}-\frac{16 l_p^2 q^4 r_h^6 \alpha ^2 E_p^{3 n}}{6-3 d}\right) r_h^{4 d}\right.\right.$$

$$+\left(-\frac{32 q^2 r_h^4 E_p^{3 n}}{4-d}-\frac{16 l_p^2 q^2 \alpha ^2 E_p^{3 n}}{d}-\frac{16 q^2 r_h^2 \left(l_p^2 \alpha ^2-2\right) E_p^{3 n}}{2-d}\right) r_h^{6 d}+\left(\frac{16 l_p^2 \alpha ^2 E_p^{3 n}}{(d-4) r_h^4}+\frac{32 E_p^{3 n}}{(d-2) r_h^2}\right) r_h^{8 d}$$
   
$$+\left(\left(\frac{16 l_p^2 q^4 \alpha ^2 E_p^{3 n}}{3 (d-2) r_h^4}+\frac{32 q^4 E_p^{3 n}}{(3 d-8) r_h^2}\right) r_h^{2 d}+\frac{32 q^6 S^{3 n} r_h^4}{4-5 (d-2)}+\frac{16 l_p^2 q^6 E_p^{3 n} \alpha ^2 r_h^2}{2-5 (d-2)}\right) r_h^{2 (d+5)}-\frac{32 q^8 E_p^{3 n} r_h^{20}}{20-7 d}$$

$$-\frac{16 l_p^2 q^8 E_p^{3 n} \alpha ^2 r_h^{18}}{18-7 d}\bigg) r_h^{-7 d}+\left(\left(\frac{10 q^6 \eta ^3 r_h^{16}}{-5 d-3 n+16}+\frac{5 l_p^2 q^6 \alpha ^2 \eta ^3 r_h^{14}}{-5 d-3 n+14}\right) r_h^{2 d}+\left(-\frac{10 q^4 \eta ^3 r_h^{10}}{10-3 (d+n)}-\frac{5 l_p^2 q^4 \alpha ^2 \eta ^3 r_h^8}{8-3 (d+n)}\right) r_h^{4 d}\right.$$

$$+\left(\frac{5 l^2 \alpha ^2 \eta ^3}{(d-3 n-4) r_h^4}+\frac{10 \eta ^3}{(d-3 n-2) r_h^2}\right) r_h^{8 d}+\left(\left(\frac{5 l^2 \alpha ^2 \eta ^3 q^4}{(3 d+3 n-8) r_h^4}+\frac{10 \eta ^3 q^4}{(3 d+3 n-10) r_h^2}\right) r_h^{2 d}\right.$$

$$+\frac{10 q^6 \eta ^3 r_h^4}{-5 d-3 n+16}+\frac{5 l_p^2 q^6 \alpha ^2 \eta ^3 r_h^2}{-5 d-3 n+14}\bigg) r_h^{2 (d+6)}-\frac{10 q^8 \eta ^3 r_h^{22}}{-7 d-3 n+22}-\frac{5 l_p^2 q^8 \alpha ^2 \eta ^3 r_h^{20}}{-7 d-3 n+20}\bigg) r_h^{-7 d-3 n}$$

$$+\left(\left(\frac{12 q^6 r_h^{16} \eta ^2 E_p^n}{-5 d-2 n+16}+\frac{6 l_p^2 q^6 r_h^{14} \alpha ^2 \eta ^2 E_p^n}{-5 d-2 n+14}\right) r_h^{2 d}+\left(-\frac{12 q^4 r_h^{10} \eta ^2 E_p^n}{-3 d-2 n+10}-\frac{6 l_p^2 q^4 r_h^8 \alpha ^2 \eta^2 E_p^n}{-3 d-2 n+8}\right) r_h^{4 d}\right.$$

$$+\left(\frac{6 l_p^2 \alpha ^2 \eta ^2 E_p^n}{(d-2 n-4) r_h^4}+\frac{12 \eta ^2 E_p^n}{(d-2 n-2) r^2}\right) r^{8 d}+\left(\left(\frac{6 l_p^2 q^4 \alpha ^2 \eta ^2 E_p^n}{(3 d+2 n-8) r^4}+\frac{12 q^4 \eta ^2 E_p^n}{(3 d+2 n-10) r^2}\right) r^{2 d}\right.$$

$$+\frac{12 q^6 E_p^n \eta ^2 r^4}{-5 d-2 n+16}+\frac{6 l_p^2 q^6 E_p^n \alpha ^2 \eta ^2 r^2}{-5 d-2 n+14}\bigg) r^{2 (d+6)}-\frac{12 q^8 E_p^n \eta ^2 r^{22}}{-7 d-2 n+22}-\frac{6 l_p^2 q^8 E_p^n \alpha ^2 \eta ^2 r^{20}}{-7 d-2 n+20}\bigg) r^{-7 d-2 n}$$

$$+\left(\left(\frac{16 q^6 r^{16} \eta  E_p^{2 n}}{-5 d-n+16}+\frac{8 l_p^2 q^6 r^{14} \alpha ^2 \eta  E_p^{2 n}}{-5 d-n+14}\right) r^{2 d}+\left(-\frac{16 q^4 r^{10} \eta  E_p^{2 n}}{-3 d-n+10}-\frac{8 l_p^2 q^4 r^8 \alpha ^2 \eta  E_p^{2 n}}{-3 d-n+8}\right) r^{4 d}\right.$$
   
$$+\left(\frac{8 l_p^2 \alpha ^2 \eta  E_p^{2 n}}{(d-n-4) r^4}+\frac{16 \eta  E_p^{2 n}}{(d-n-2) r^2}\right) r^{8 d}+\left(\left(\frac{8 l_p^2 q^4 \alpha ^2 \eta  E_p^{2 n}}{(3 d+n-8) r^4}+\frac{16 q^4 \eta  E_p^{2 n}}{(3 d+n-10) r^2}\right) r^{2 d}\right.$$

\begin{equation}\label{ah9.equn35}
+\frac{16 q^6 E_p^{2 n} \eta  r^4}{-5 d-n+16}+\frac{8 l_p^2 q^6 E_p^{2 n} \alpha ^2 \eta  r^2}{-5 d-n+14}\bigg) r^{2 (d+6)}-\frac{16 q^8 E_p^{2 n} \eta  r^{22}}{-7 d-n+22}-\frac{8 l_p^2 q^8 E_p^{2 n} \alpha ^2 \eta  r^{20}}{-7 d-n+20}\bigg) r^{-7 d-n}\bigg)\pi.
\end{equation} 
\end{document}